\documentclass{svmult}

\usepackage{amsmath,amssymb} 
\usepackage{makeidx}         
\usepackage{graphicx}        
\usepackage{multicol}        
\usepackage[bottom]{footmisc}

\newcommand{\Teilprojekt}{}  

\begin{document}

\renewcommand{\Teilprojekt}{A7}

\title*{Computer Simulation of Particle Suspensions}

\author{Jens Harting\inst{1} \and Martin Hecht\inst{1} \and Hans J. Herrmann\inst{2} \and Sean McNamara\inst{1}}
\authorrunning{Harting, Hecht, Herrmann, McNamara}

\institute{Institute for Computational Physics,  University of Stuttgart,
Pfaffenwaldring 27, 70569 Stuttgart, Germany \\
\texttt{jens@icp.uni-stuttgart.de}  \\
\texttt{Martin.Hecht@icp.uni-stuttgart.de} \\
\texttt{S.McNamara@icp.uni-stuttgart.de}\and 
Institute for Building Materials,
ETH H\"{o}nggerberg,
HIF E 12,
8093 Z\"{u}rich, Switzerland \\
\texttt{hans@icp.uni-stuttgart.de} }
%
%

\maketitle

\begin{abstract}
Particle suspensions are ubiquitous in our daily life, but are not well
understood due to their complexity. During the last twenty years, various
simulation methods have been developed in order to model these systems.
Due to varying properties of the solved particles and the solvents, one
has to choose the simulation method properly in order to use the available
compute resources most effectively with resolving the system as well as
needed. Various techniques for the simulation of particle suspensions have
been implemented at the Institute for Computational Physics allowing us to
study the properties of clay-like systems, where Brownian motion is
important, more macroscopic particles like glass spheres or fibers solved
in liquids, or even the pneumatic transport of powders in pipes.  In this
paper we will present the various methods we applied and developed and
discuss their individual advantages. 
\end{abstract}

\index{particle suspensions}
\index{molecular dynamics}
\index{stochastic rotation dynamics}
\index{lattice Boltzmann method}
\index{colloids}

\section{Introduction} \label{A7:Introduction}
Adding a fluid to a dry granulate causes the behavior of the mixture to
change dramatically and a host of unexpected phenomena arises. A good
example can be studied by anyone on the beach: whereas it is impossible to
build a sand castle from dry sand, once just a little bit
of water has been stirred into the sand, one can shape the resulting
material almost arbitrarily into surprisingly complex arrangements. Adding
even more fluid might result in the material loosing this stability.
If we stir such a mixture, it behaves like a liquid of increased density. 
Other very common particle-fluid mixtures are ubiquitous in our daily
life and include the cacao drink which keeps separating into its
constituents, tooth paste and wall paint which are mixtures of finely
ground solid ingredients in fluids or blood which is made up of red and
white blood cells suspended in a solvent. An extreme example is the sand
on the beach which can be blown away by the wind. 
It is important for industrial applications to obtain a detailed
knowledge of those systems in order to optimize production processes or to
prevent accidents.

Long-range fluid-mediated hydrodynamic interactions often dictate the
behavior of particle-fluid mixtures. The majority of analytical results
for the particle scale behavior of suspensions has been obtained in the
regime of vanishing Reynolds numbers (viscous flow). For large systems,
scientists aim at an average, continuum
description of the large-scale behavior. However, this requires
time-consuming and sometimes very difficult experimental measurements
of phenomenological quantities such as the mean settling speed of a
suspension, the stress contributions in the system of the individual
components (solid and fluid) as functions of, e.g., the solid volume
fraction of the constituents.

Computer simulation methods are indispensable for many-particle systems,
for the inclusion of inertia effects (Reynolds numbers $> 1$) and Brownian
motion (Peclet number of order $1$). These systems often contain a large
number of important time scales which differ by many orders of magnitude,
but nevertheless have to be resolved by the simulation, leading to a large
numerical effort. However, simulations have the potential to increase our
knowledge of elementary processes and to enable us to find the
aforementioned relations from simulations instead of experiments.

Various simulation methods have been developed to simulate particle-fluid
mixtures. All of them have their inherent strengths but also some
disadvantages. For example, simplified Brownian Dynamics does not contain
long-ranged hydrodynamic interactions among particles at
all~\cite{A7:Huetter00}. Brownian Dynamics with full hydrodynamic
interactions utilizes a mobility matrix which is based on tensor
approximations which are exact in the limit of zero Reynolds number and
zero particle volume fraction\;\cite{A7:Petera99, A7:Ahlrichs01}. However,
the computational effort scales with the cube of the particle number due
to the inversion of matrices.  Pair-Drag simulations have been proposed by
Silbert et al.~\cite{A7:Silbert97a}, and include hydrodynamic interactions
in an approximative way. They have focused on suspensions with high
densities (up to $50\,\%$) of uncharged spherical colloidal particles.
Stokesian Dynamics has been presented by Bossis and Brady in the 80s and
applied by many authors
\cite{A7:bossis84a,A7:sierou01a,A7:phung96a,A7:brady:88}. For example,
Melrose and Ball have performed detailed studies of shear thickening
colloids using Stokesian Dynamics simulations
\cite{A7:melrose04e,A7:melrose04d}. However, this method is limited to
Reynolds numbers close to zero and the computational effort is very high
for dynamical simulations. Even with today's powerful computers it is not
possible to study the dynamics of more than a few hundred particles. The
method is still widely used due to its physical motivation and its
robustness, but is complicated to code.  Boundary-element methods are more
flexible than Stokesian dynamics and can also be used to simulate
non-spherical or deformable particles, but the computational effort is
even higher~\cite{A7:loewenberg96,A7:Ladd01}.

All these methods assume that hydrodynamic interactions are fully
developed and that the dynamics of the fluid and of the solved particles
can be treated as fully separated. In reality, this is not the case.
Hydrodynamic interactions are time dependent due to local stresses at the
fluid-particle interfaces. A number of more recent methods attempt to
describe the time dependent long-range hydrodynamics properly with the
computational effort scaling linearly with the number of particles. These
include recent mesoscopic methods like dissipative particle dynamics
\cite{A7:espanol-warren,A7:espanol,A7:boek-coveney-lekkerkerker-vanderschoot},
the lattice-Boltzmann method
\cite{A7:chen98a,A7:Ladd01,A7:Ladd94,A7:Ladd94b,A7:nguyen02,A7:jens-komnik-herrmann:2004},
or stochastic rotation dynamics \cite{A7:Malev99,
A7:Malev00,A7:Hecht05,A7:HHBRH06}. However, for small Reynolds numbers,
the computational gain of these methods is lost due to the additional
effort needed to describe the motion of the fluid.
Finite element or finite difference methods need a proper meshing of the computational domain
which is not trivial for complicated boundary conditions as in the case of
dense suspensions. Therefore, many authors only simulated a limited number
of static configurations rather than the full dynamics of the system.
Advances in remeshing techniques as well as more powerful computers have
allowed to overcome these problems. Also, in order to avoid remeshing at
all, uniform grids can be used \cite{A7:Fogelson88,A7:Hoefler99,A7:Hoefler00-a}.
These methods are flexible and robust. They can properly treat
non-Newtonian effects and incorporate inertia, but are complicated to
code.

For a more detailed description of the simulation methods, experiments or
theoretical approaches not addressed in this paper, the reader is referred
to one of the various books on colloid science
\cite{A7:Mahanty76, A7:Lagaly97,A7:Shaw92,A7:Morrison02,A7:Schmitz93,A7:Hunter01}.

The remainder of this paper focuses on three different simulation
techniques which have recently been applied to particle-laden flows in our group.
First, we introduce a method developed by Malevanets and Kapral to model a
solvent with thermal fluctuations. This approach is used to study the
properties of claylike colloids \cite{A7:Hecht05,A7:HHBRH06}. For larger
particles, thermal fluctuations are undesirable. Here, the lattice
Boltzmann method and its extension to particle suspension is a very good
candidate to study the dynamics of glass spheres in a sugar solution
\cite{A7:jens-komnik-herrmann:2004,A7:jens-harvey-chin-venturoli-coveney:2005}.
The method is easy to code and has been applied to suspensions of
spherical and non-spherical particles by various authors.
If the particles are very massive and the density of the fluid is very
low, a full hydrodynamic treatment of the solvent is not needed anymore.
In the last chapter we describe an algorithm based on a coarse-grained
description of the fluid, so that it is resolved on a length scale larger
than the particles. Much larger systems can be treated this way, but the
coarse-graining is justified only in certain situations. As an example we
model the pneumatic transport of a powder in a pipe which is a common
process in many industrial applications \cite{A7:MCN0003,A7:Strauss1,A7:Strauss2}.

A more computational demanding and not as easy to code method is a Navier
Stokes solver for the fluid which is coupled to the particles. The method
has been successfully applied to the simulation of sedimentation processes
of spherical or non-spherical particles and profits from its well
established physical background and long standing experience with similar
fluid solvers in engineering disciplines \cite{A7:Wachmann98,A7:Hoefler99,A7:Hoefler00-a,A7:Kuusela01,A7:Fonseca04a,A7:Fonseca04b}.
These methods have a long standing history in our group, but have been
described in detail elsewhere and will not be covered in this paper.

\section{Simulation of Claylike Colloids: Stochastic Rotation
Dynamics}\label{A7:sec:SRD}
\index{peloids}
Dense suspensions of small strongly interacting particles are complex
systems, which are rarely understood on the microscopic level.
We investigate properties of dense
suspensions and sediments of small spherical Al$_2$O$_3$ particles 
by means of a combined Molecular Dynamics (MD) and Stochastic
Rotation Dynamics (SRD) simulation. 
Stochastic Rotation Dynamics is a simulation method developed 
by Malevanets and Kapral~\cite{A7:Malev99, A7:Malev00} for a 
fluctuating fluid. The work this chapter is dealing with is
presented in more detail in references~\cite{A7:Hecht05,A7:HHBRH06}.

We simulate claylike colloids, for which in many cases the attractive Van-der-Waals forces are relevant.
They are often called ``peloids'' (Greek: clay-like). The colloidal particles 
have diameters in the range of some nm up to some $\umu$m. The term ``\emph{peloid}''
originally comes from soil mechanics, but particles of this size are also important in many
engineering processes. Our model systems of Al$_2$O$_3$-particles of about
half a $\umu$m in diameter 
suspended in water are often used ceramics and play an important role in technical processes.
In soil mechanics~\cite{A7:Richter03} and ceramics science~\cite{A7:Oberacker01},
questions on the shear viscosity and compressibility
as well as on porosity of the microscopic structure which is formed by the particles,
arise~\cite{A7:Wang99,A7:Lewis00}. In both areas, usually high volume fractions ($\Phi > 20\%$) are of interest.
The mechanical properties of these suspensions
are difficult to understand. Apart from the attractive forces, electrostatic repulsion strongly
determines the properties of the suspension. Depending on the surface potential, 
one can either observe formation of clusters or the particles are stabilized in suspension and
do sediment only very slowly. The surface potential can be adjusted by the $p$H-value of the solvent.
Within Debye-H\"uckel theory one can derive a so-called $2pK$ charge regulation model which relates
the simulation parameters with the $p$H-value and ionic strength $I$ adjusted in the experiment.
In addition to the static interactions hydrodynamic effects are also important for a complete 
description of the suspension. Since typical Peclet numbers are of order one in our system, 
Brownian motion cannot be neglected.

\subsection{Molecular Dynamics (MD): Simulation of the Suspended Particles}
\label{A7:SRD-MD}
\index{molecular dynamics}
We study colloidal particles, composing the solid fraction, suspended in a fluid
solvent. The colloidal particles are simulated with molecular dynamics (MD), whereas
the solvent is modeled with stochastic rotation dynamics (SRD) as described in 
Sect.\ \ref{A7:secm_srd}.
\\
In the MD part of our simulation we include effective electrostatic interactions and
van der Waals attraction, a lubrication force and Hertzian contact forces.
In order to correctly model the statics and dynamics when approaching
stationary states, realistic potentials are needed.
The interaction between the particles is described by DLVO theory~\cite{A7:Huetter00,A7:Russel95,A7:Lewis00}.
If the colloidal particles are
suspended in a solvent, typically water, ions move into solution, whereas their counter ions
remain in the particle due to a different resolvability. Thus, the colloidal particle carries
a charge. The ions in solution are attracted by the charge on the particles and form the
electric double layer. It has been shown\,(see \cite{A7:Russel95}), that the resulting electrostatic
interaction between two of these particles can be described by an exponentially screened Coulomb potential
\begin{equation}
  V_{\mathrm{Coul}} =
  \pi \varepsilon_r \varepsilon_0
  \left[ \frac{2+\kappa d}{1+\kappa d}\cdot\frac{4 k_{\mathrm{B}} T}{z e}
         \tanh\left( \frac{z e \zeta}{4 k_{\mathrm{B}} T} \right)
  \right]^2 
   \times \frac{d^2}{r} \exp( - \kappa [r - d])\;,
 \label{A7:eqm_VCoul}
\end{equation}
\index{DLVO theory}
where $d$ denotes the particle diameter and $r$ is the distance between the 
particle centers. $e$ is the elementary charge, $T$ the temperature, 
$k_{\mathrm{B}}$ the Boltzmann constant, and $z$ is the valency of the ions of 
added salt. Within DLVO theory one assumes linear screening, mainly by one 
species of ions with valency $z$ (e.g. $z=+1$ for NH$_4^+$). The first fraction
in \eqref{A7:eqm_VCoul} is a correction to the original DLVO potential, which takes
the surface curvature into account and is valid for spherical particles~\cite{A7:Trizac02}.

The effective surface potential $\zeta$ is the electrostatic potential 
at the border between the diffuse layer and the compact layer, it may therefore be 
identified with \emph{the} $\zeta$-potential. It includes the effect of the bare charge 
of the colloidal particle itself, as well as the charge of the ions in the Stern layer,
where the ions are bound permanently to the colloidal particle. In other words, 
DLVO theory uses a renormalized surface charge. This charge can be related to the
$p$H value of the solvent within Debye-H\"{u}ckel theory~\cite{A7:HHBRH06}.

$\varepsilon_0$ is the permittivity of the vacuum, $\varepsilon_r$ the relative dielectric
constant of the solvent. $\kappa$ is the inverse Debye length defined by 
$\kappa^2 = 8\pi\ell_BI$, with the 
ionic strength $I$ and the Bjerrum length $\ell_B$. 
We use 81 for water, which implies  $\ell_{B} = 7\,$\AA.

\begin{figure}
\begin{center}
\includegraphics[scale=0.3]{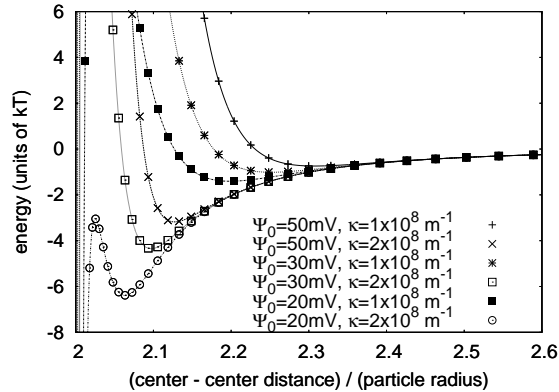}
\end{center}
\caption{DLVO potentials for Al$_2$O$_3$ spheres of $R=0.5\,\umu$m diameter suspended in water.
These are typical potentials used for our simulations as described below. The primary minimum
at $d/R = 2.0$ is not reproduced correctly by the DLVO theory. It has to be modeled separately.
In most of our cases the existence of the secondary minimum determines the properties of the simulated
system}
\label{A7:figm_Potentials}
\end{figure}

The Coulomb term of the DLVO potential competes with the attractive van der Waals term
\begin{equation}
  V_{\mathrm{VdW}} = - \frac{A_{\mathrm{H}}}{12} 
     \left[ \frac{d^2}{r^2 - d^2} + \frac{d^2}{r^2} \right. 
             \;\left. + 2 \ln\left(\frac{r^2 - d^2}{r^2}\right) \right]\;.
\end{equation}
$A_{\mathrm{H}}=4.76\cdot 10^{-20}\,\mathrm{J}$ is the Hamaker constant~\cite{A7:Huetter99} 
which involves the polarizability of the particles. It is kept constant in our simulations 
since it only depends on the material of the particles and on the solvent. 
The DLVO potentials are plotted in Fig.\ \ref{A7:figm_Potentials} for six
typical examples with different depth of the secondary minimum. The primary minimum has to
be modeled separately, as discussed below.
Long range hydrodynamic interactions are taken into account in the simulation for
the fluid as described below. This can only reproduce interactions correctly down
to a certain length scale. On shorter distances, a lubrication force has to be
introduced explicitly in the molecular dynamics simulation.
The most dominant mode, the so-called squeezing mode, is an additional force
\begin{eqnarray}
  \label{A7:eqm_FLub}
  \mathbf{F}_{\mathrm{lub}} &=& -(\mathbf{v}_{\mathrm{rel}},\mathbf{\hat{r}})\mathbf{\hat{r}}
    \frac{6 \pi \eta r_{\mathrm{red}}^2}{r - r_1 -r_2}\;, \\
  \mathrm{with\quad}r_{\mathrm{red}} &=& \frac{r_1 r_2}{r_1+r_2}
\end{eqnarray}
between two spheres with radii $r_1$, $r_2$ and the relative velocity $\mathbf{v}_{\mathrm{rel}}$.
$\eta$ is the dynamic viscosity of the fluid. In contrast to the DLVO potentials
the lubrication force is a dissipative force. When two particles approach each
other very closely, this force becomes very large. To ensure numerical stability
of the simulation, one has to limit $\mathbf{F}_{\mathrm{lub}}$. We do this
by introducing a small cutoff radius $r_{\mathrm{sc}}$. Instead of calculating 
$\mathbf{F}_{\mathrm{lub}}(r)$ we take the value for 
$\mathbf{F}_{\mathrm{lub}}(r+r_{\mathrm{sc}})$. In addition, since the force decays for 
large particle distances, we can introduce a large cutoff radius $r_{\mathrm{lc}}$ for which
we assume $\mathbf{F}_{\mathrm{lub}}(r) \equiv 0$ if $r > r_{\mathrm{lc}}$.
As the intention of $\mathbf{F}_{\mathrm{lub}}$ is to correct the 
finite resolution of the fluid simulation, $r_{\mathrm{sc}}$ and $r_{\mathrm{lc}}$
have to be adjusted in a way that the dynamic properties, i.e., the viscosity
of a simulated particle suspension with weak DLVO interactions fit the measurements.
It turns out that $r_{\mathrm{sc}} = 1.05 (r_1 + r_2)$ and $r_{\mathrm{lc}} = 2.5 (r_1 + r_2)$
work best. 
\\
To avoid that the particles penetrate each other, one needs a repulsive force depending
on their overlap. We are using a Hertz force described by the potential
\begin{equation}
  V_{\mathrm{Hertz}} = K (d-r)^{5/2}  \quad  \mathrm{if}  \quad r<d\;,
\end{equation}
where $K$ could be expressed by the elastic modulus of Al$_2$O$_3$. This would determine
the simulation time step, but to keep the computational effort relatively small, we determine
the time step using the DLVO-potentials as described later on and then choose a value for
$K$. Two aspects have to be considered: $K$ has to be big enough so that the particles
do not penetrate each other by more than approximately $10\%$ and it may not be too big, so that
numerical errors are kept small, which is the case when the collision time is resolved
with about 20 time steps. Otherwise total energy and momentum are not conserved very well
in the collision.
\\
The Hertz force also contains a damping term in normal direction,
\begin{equation}
  \mathbf{F}_{\mathrm{Damp}} =  -(\mathbf{v}_{\mathrm{rel}},\mathbf{\hat{r}})\mathbf{\hat{r}}
    \beta  \sqrt{r - r_1 - r_2}\;,
\end{equation}
with a damping constant $\beta$
and for the transverse direction a viscous friction proportional to the relative velocity of
the particle surfaces is applied.
\\
Since DLVO theory contains the assumption of linear polarizability, it holds only for
large distances, i.e., the singularity when the two spheres touch does not exist in reality.
Nevertheless, there \emph{is} an energy minimum about $30\,k_{\mathrm{B}}T$ deep, so that
particles which come that close would very rarely become free again. To obtain numerical
stability of our simulation, we model this minimum by a parabolic potential, some
$k_{\mathrm{B}}T$ deep (e.g. $6\,k_{\mathrm{B}}T$).
The depth of the minimum in our model is much less than in reality, but the probability
for particles to be trapped in the minimum has to be kept low enough so that only few
of them might escape during simulation time.

\index{velocity verlet}
For the integration of the translational motion we utilize a velocity
Verlet algorithm~\cite{A7:Allen87}
to update the velocity and position of particle $i$ according to the equations
\begin{eqnarray}
 {\bf x}_i(t+\delta t) &=& {\bf x}_i (t) + \delta t \,{\bf v}_i(t)+\delta
t^2\,\frac{F_i(t)}{m}\;, \\
 {\bf v}_i(t+\delta t) &=& {\bf v}_i (t) + \delta t \,\frac{F_i(t) +
F_i(t+\delta t)}{2 m}\; .
\end{eqnarray}
For the rotation, a simple Euler algorithm is applied:
\begin{eqnarray}
  {\bf\omega}_i(t+\delta t)&=&{\bf\omega}_i(t) + \delta t \,{\bf T}_i\;, \\
  {\bf\vartheta}_i(t+\delta t)&=&{\bf\vartheta}_i(t) + F({\bf\vartheta}_i,
{\bf\omega}_i, \delta t)\;,
\end{eqnarray}
where ${\bf\omega}_i(t)$ is the angular velocity of particle $i$ at time $t$, ${\bf T}_i$ is
the torque exerted by non central forces on the particle $i$, ${\bf\vartheta}_i(t)$ is the orientation
of particle $i$ at time $t$, expressed by a quaternion, and $F({\bf\vartheta}_i, {\bf\omega}_i, \delta t)$
gives the evolution of ${\bf\vartheta}_i$ of particle $i$ rotating with the angular velocity ${\bf\omega}_i(t)$
at time $t$.
\\
The concept of quaternions~\cite{A7:Allen87} is often used to calculate rotational motions in simulations,
because the Euler angles and rotation matrices can easily be derived from quaternions. Using Euler angles
to describe the orientation would give rise to singularities for the two orientations
with $\vartheta = \pm 90^{\circ}$. The numerical problems related to this fact and the relatively high
computational effort of a matrix inversion can be avoided using quaternions.

\subsection{Stochastic Rotation Dynamics (SRD): Simulation of the Fluid}
\noindent
\label{A7:secm_srd}
The Stochastic Rotation Dynamics method (SRD) introduced by Malevanets and Kapral \cite{A7:Malev99, A7:Malev00}
 is a promising tool for a coarse-grained description of
a fluctuating solvent, in particular for colloidal and polymer suspensions. The method is also
known as ``Real-coded Lattice Gas'' \cite{A7:Inoue02} or as ``multi-particle-collision dynamics'' (MPCD)
\cite{A7:Gompper04}.
\index{real-coded lattice gas}
\index{Malevanets-Kapral method}
It can be seen as a ``hydrodynamic heat bath'', whose details are not fully resolved but which
provides the correct hydrodynamic interaction among embedded particles~\cite{A7:Lamura01}.
SRD is especially well suited for flow problems with Peclet numbers of 
order one and Reynolds numbers on the particle scale between 0.05 and 20 
for ensembles of many particles.
The method is based on so-called fluid particles with continuous positions and velocities.
Each time step is composed of two simple steps: One streaming step and one interaction step.
In the streaming step the positions of the fluid particles are updated as in the
Euler integration scheme known from Molecular Dynamics simulations: 
\begin{equation}
\label{A7:eqm_move}
{\bf r}_i(t+\tau)={\bf r}_i(t)+\tau\;{\bf v}_i(t)i\;,
\end{equation}
where ${\bf r}_i(t)$ denotes the position of the particle $i$ at time $t$, 
${\bf v}_i(t)$ its velocity at time $t$ and $\tau$ is the time step used for
the SRD simulation. 
After updating the positions of all fluid particles they interact collectively in 
an interaction step which is constructed to preserve momentum, energy and particle number.
The fluid particles are sorted into cubic cells of a regular lattice and only the  
particles within the same cell are involved in the interaction step. First, their
mean velocity ${\bf u}_j(t')=\frac{1}{N_j(t')}\sum^{N_j(t')}_{i=1} {\bf v}_i(t)$
is calculated, where ${\bf u}_j(t')$ denotes the mean velocity of cell $j$ containing
$N_j(t')$ fluid particles at time $t'=t+\tau$. Then, the velocities of each fluid particle in 
cell $j$ are updated as:
\begin{equation}
\label{A7:eqm_rotate}
{\bf v}_i(t+\tau) = {\bf u}_j(t')+{\bf \Omega}_j(t') \cdot [{\bf
v}_i(t)-{\bf u}_j(t')]\;.
\end{equation}
${\bf \Omega}_j(t')$ is a rotation matrix, which is independently chosen randomly
for each time step and each cell. We use rotations about one of the coordinate axes by 
an angle $\pm\alpha$, with $\alpha$ fixed. This has been suggested by M. Strau{\ss} 
in~\cite{A7:Ihle03c}.
The coordinate axis as well as the sign of the rotation are chosen by random, 
resulting in six possible rotation matrices. The mean velocity ${\bf u}_j(t)$ in 
the cell $j$ can be seen as streaming velocity of the fluid at the position of the cell 
$j$ at the time $t$, whereas the difference $[{\bf v}_i(t)-{\bf u}_j(t')]$ entering 
the interaction step can be interpreted as a contribution to the thermal 
fluctuations.

In order to remove low temperature anomalies and to achieve exact Galilean-invariance, we use
a modification of the original algorithm~\cite{A7:Ihle01}:
all particles are shifted by the {\it same}
random vector with components in the interval $[-a/2,a/2]$ before the collision step.
Particles are then shifted back by the same amount after the collision.
The random vectors of consecutive iterations are uncorrelated.
Ihle and Kroll have discussed in~\cite{A7:Ihle03a, A7:Ihle03b} why this simple procedure works and shown
that it leads to transport coefficients independent of an imposed homogeneous flow field.
In~\cite{A7:Ihle04} and~\cite{A7:Yeomans03} analytical calculations of the transport
coefficient of this method are presented.

Two different methods to couple the SRD and the MD simulation have been
introduced in the literature. Inoue et al. proposed a way to implement no
slip boundary conditions on the particle surface~\cite{A7:Inoue02},
whereas Falck et al.~\cite{A7:Falck04} have developed 
a ``more coarse grained'' method we describe shortly in the following section.

\subsection{Coupling of the MD and the SRD Simulation Part}
\noindent
To couple the two parts of the simulation, MD on the one hand and SRD on the other one, 
the colloidal particles are sorted into
the SRD boxes and their velocities are included in the rotation step. 
This technique has been used to model 
protein chains suspended in a liquid~\cite{A7:Falck04, A7:Gompper04b}. Since the mass
of the fluid particles is much smaller than the mass of the colloidal particles, 
one has to use the mass of each particle---colloidal or fluid particle---as a weight factor when 
calculating the mean velocity 
\begin{eqnarray}
\label{A7:eqm_rotateMD}
{\bf u}_j(t')&=& \frac{1}{M_j(t')}\sum\limits^{N_j(t')}_{i=1}{\bf v}_i(t)
m_i\;,
\\
\label{A7:eqm_rotateMD2}
\mathrm{with}\qquad M_j(t')&=&\sum^{N_j(t')}_{i=1}m_i\;,
\end{eqnarray}
where we sum over all colloidal and fluid particles in the cell, so that $N_j(t')$ is 
the total number of both particles together. $m_k$ is the mass of the
particle with index $i$ and therefore $M_j(t')$ gives the total mass contained 
in cell $j$ at the time $t'=t+\tau$. 

\subsection{Results}

\subsubsection{Phase Diagram}

\begin{figure}
\begin{center}
\includegraphics[scale=0.3]{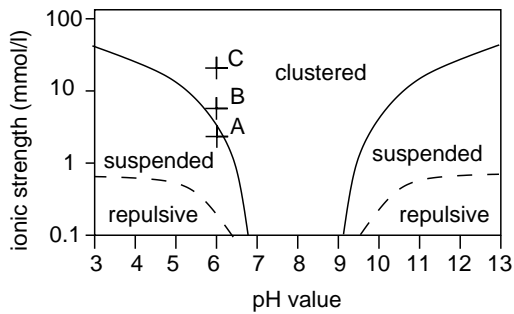}
\end{center}
\caption{Schematic phase diagram for volume fraction $\Phi=35\%$ in
terms of $p$H-value and ionic strength involving three different
phases: a clustering regime due to van der Waals attraction, stable  
suspensions where the charge of the colloidal particles prevents
clustering, and a repulsive structure for further increased electrostatic repulsion.
This work concentrates on state $A$ ($p\mathrm{H}=6$, $I=3\,$mmol/l) in the
suspended phase, state $B$ ($p\mathrm{H}=6$, $I=7\,$mmol/l) close to the phase border but 
already in the clustered phase, and state $C$ ($p\mathrm{H}=6$,
$I=25\,$mmol/l) in the clustered phase \cite{A7:HHBRH06}
}
\label{A7:figm_phasediag}
\end{figure}

Depending on the experimental conditions, one can obtain three different phases: A clustered
region, a suspended phase, and a repulsive structure. These phases can be reproduced in the 
simulations and we can quantitatively relate interaction potentials to certain experimental 
conditions. 
A schematic picture of the phase diagram is shown in Fig.\ \ref{A7:figm_phasediag}. 
Close to the isoelectric point ($p\mathrm{H}=8.7$), the particles form clusters for all 
ionic strengths since they are not charged. At lower or higher $p$H values
one can prepare a stable suspension for low ionic strengths because of the charge, which
is carried by the colloidal particles. At even more extreme $p$H values, one can obtain
a repulsive structure due to very strong electrostatic potentials (up to $\zeta = 170\,$mV 
for $p\mathrm{H} = 4$ and $I = 1\,$mmol/l, according to our model). The repulsive structure
is characterized by an increased shear viscosity. 
In the following we focus on three states: State $A$ ($p\mathrm{H}=6$, $I=3\,$mmol/l) is in the
suspended phase, state $B$ ($p\mathrm{H}=6$, $I=7\,$mmol/l) is a point already in the clustered 
phase but still close to the phase border, and state $C$ ($p\mathrm{H}=6$, $I=25\,$mmol/l) is 
located well in the clustered phase.

\begin{figure}
\begin{center}
\parbox{0.4\linewidth}{\includegraphics[scale=0.2]{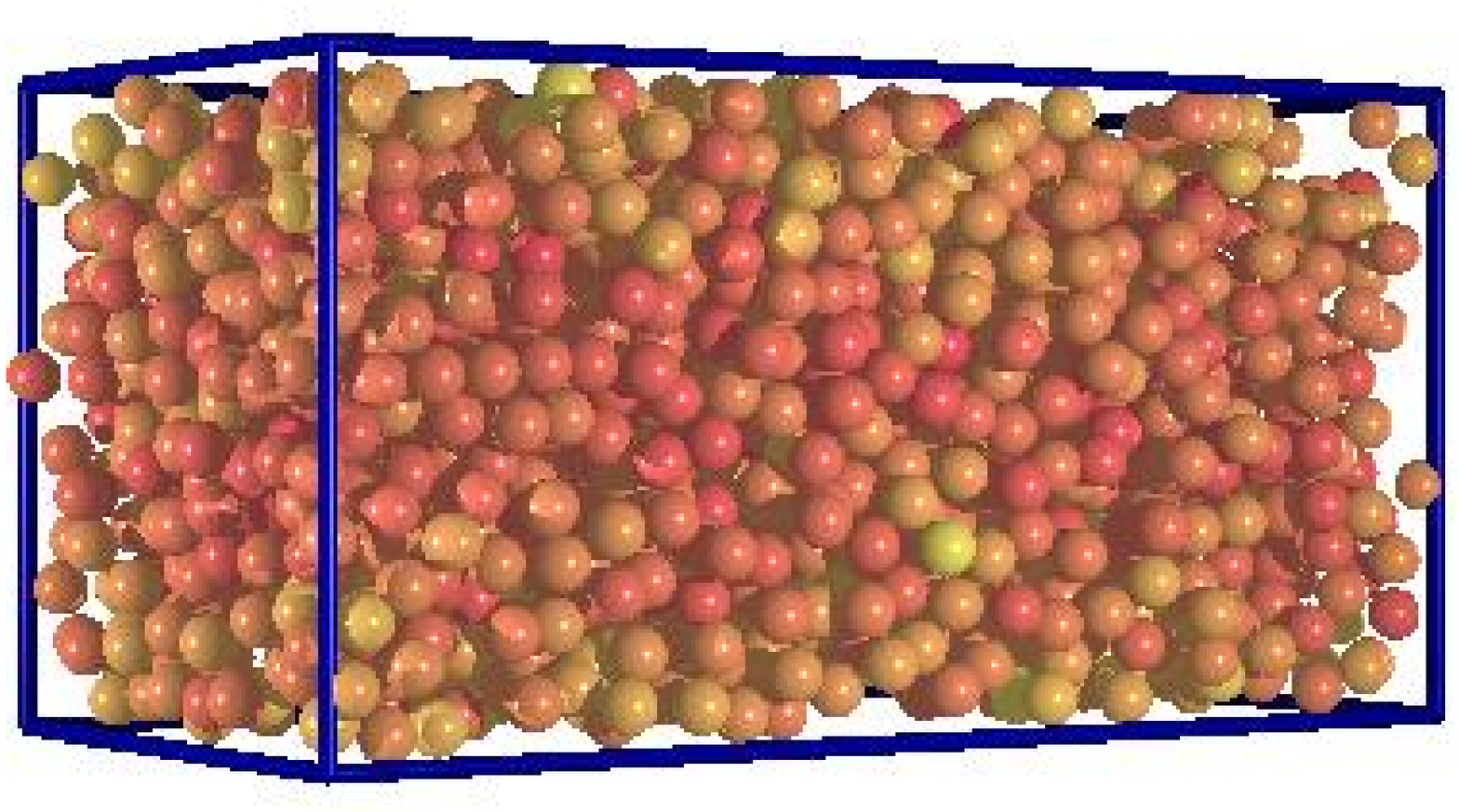} \\ {\bf
a) suspended case}}
\parbox{0.4\linewidth}{\includegraphics[scale=0.2]{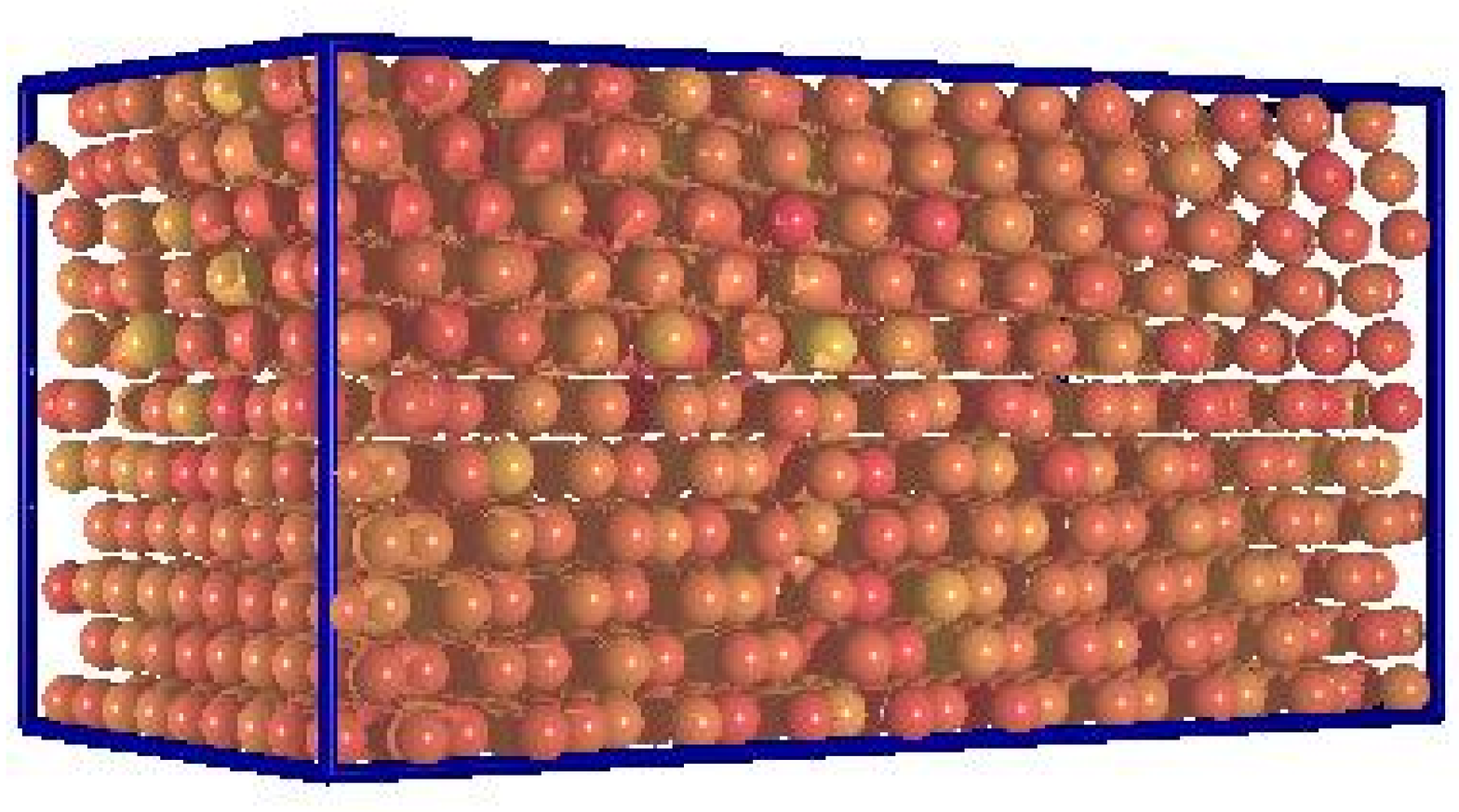} \\ {\bf
b) layer formation}}
\parbox{0.4\linewidth}{\includegraphics[scale=0.2]{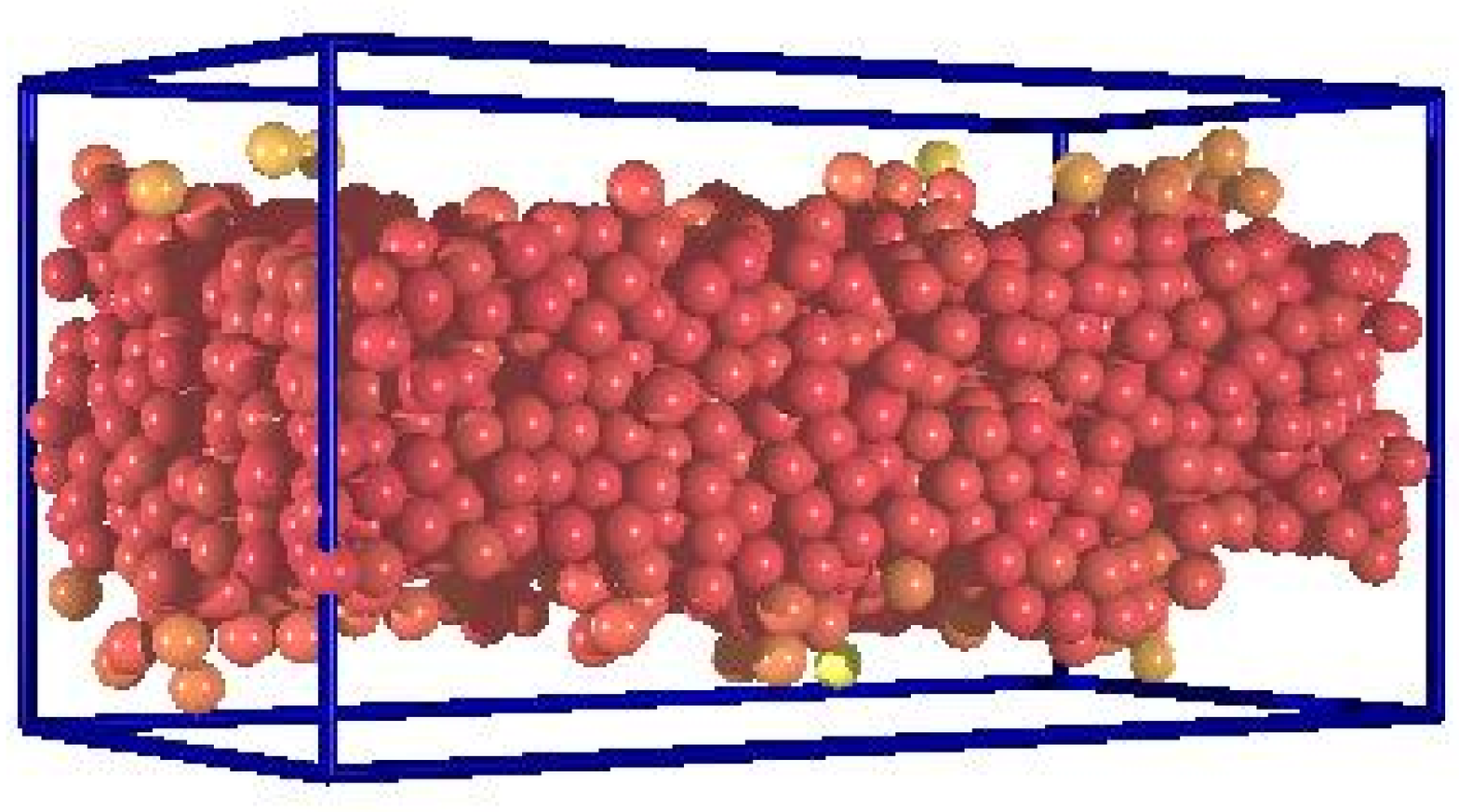} \\ {\bf
c) central cluster}}
\parbox{0.4\linewidth}{\includegraphics[scale=0.2]{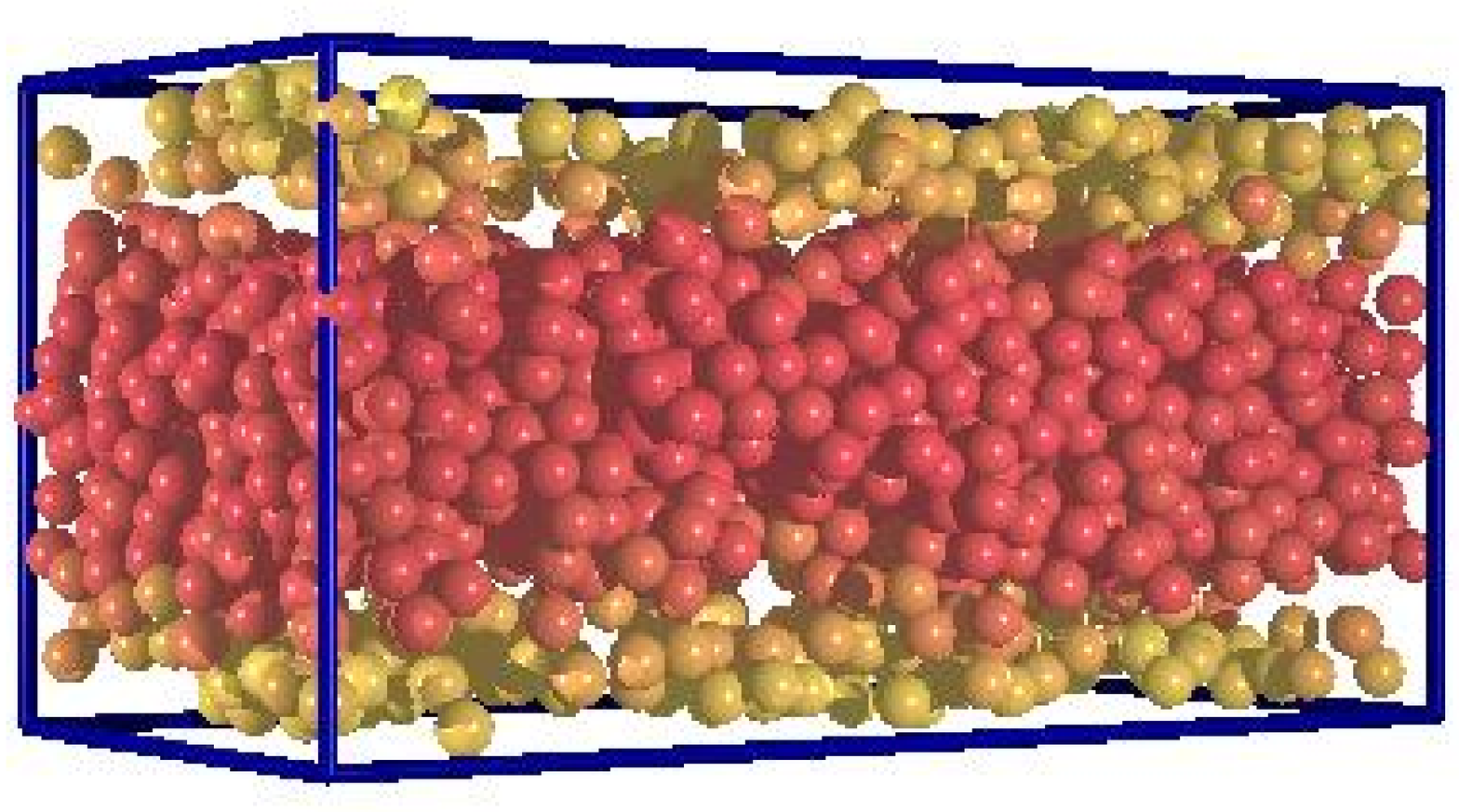} \\ {\bf
d) plug flow}}
\end{center}
\caption{Images of four different cases. For better visibility we have chosen smaller systems than we 
usually use for the calculation of the viscosity. The colors denote velocities: Dark particles are
slow, bright ones move fast. The potentials do not correspond exactly to the
cases $A$--$C$ in Fig.\ \ref{A7:figm_phasediag}, but they show qualitatively the differences between the 
different states: 
\textbf{a)} Suspension like in state $A$, at low shear rates. 
\textbf{b)} Layer formation, which occurs in the repulsive regime, but also in the suspension (state $A$) at high shear rates.
\textbf{c)} Strong clustering, like in state $C$, so that the single cluster in the simulation is deformed. 
\textbf{d)} Weak clustering close to the phase border like in state $B$, where the cluster can be broken into pieces, which follow the flow of the fluid (plug flow)
}
\label{A7:figm_Snapshots}
\end{figure}

Some typical examples for the different phases are shown in Figs.\ \ref{A7:figm_Snapshots}a)--d).
These examples are meant to be only illustrative and do not correspond exactly to the
cases $A$--$C$ in Fig.\ \ref{A7:figm_phasediag} denoted by uppercase letters. 
In the suspended case (a), the particles are mainly coupled 
by hydrodynamic interactions. One can find a linear velocity profile and a slight shear thinning. 
If one increases  the shear rate $\dot\gamma>500$/s, the particles arrange in layers. 
The same can be observed if the Debye-screening length of the electrostatic potential 
is increased (b), which means that the solvent contains less ions ($I < 0.3\,$mmol/l) to screen 
the particle charges. On the other hand, if one increases the salt 
concentration, electrostatic repulsion is screened even more and attractive van der Waals
interaction becomes dominant ($I > 4\,$mmol/l). Then the particles form clusters (c), and 
viscosity rises. A special case, called ``plug flow'', can be observed for high shear rates, 
where it is possible to tear the clusters apart and smaller parts of them follow with the 
flow of the solvent (d). This happens in our simulations for $I = 25\,$mmol/l (state $C$) 
at a shear rate of $\dot\gamma > 500$/s. However, as long as there are only one or two big 
clusters in the system, it is too small to expect quantitative agreement with experiments.
In these cases we have to focus on state $B$ ($I = 7\,$mmol/l) close to the phase border.

We restrict ourselves to the region around $p\mathrm{H}=6$ where we find
the phase border between the suspended region and the clustered regime at about $I=4\,$mmol/l
in the simulations as well as in the experiments. Also the shear rate dependence of the
viscosity is comparable in simulations and experiments as discussed below.

\subsubsection{Shear Profile and Shear Viscosity}
\begin{figure}
\begin{center}
\includegraphics[scale=0.3]{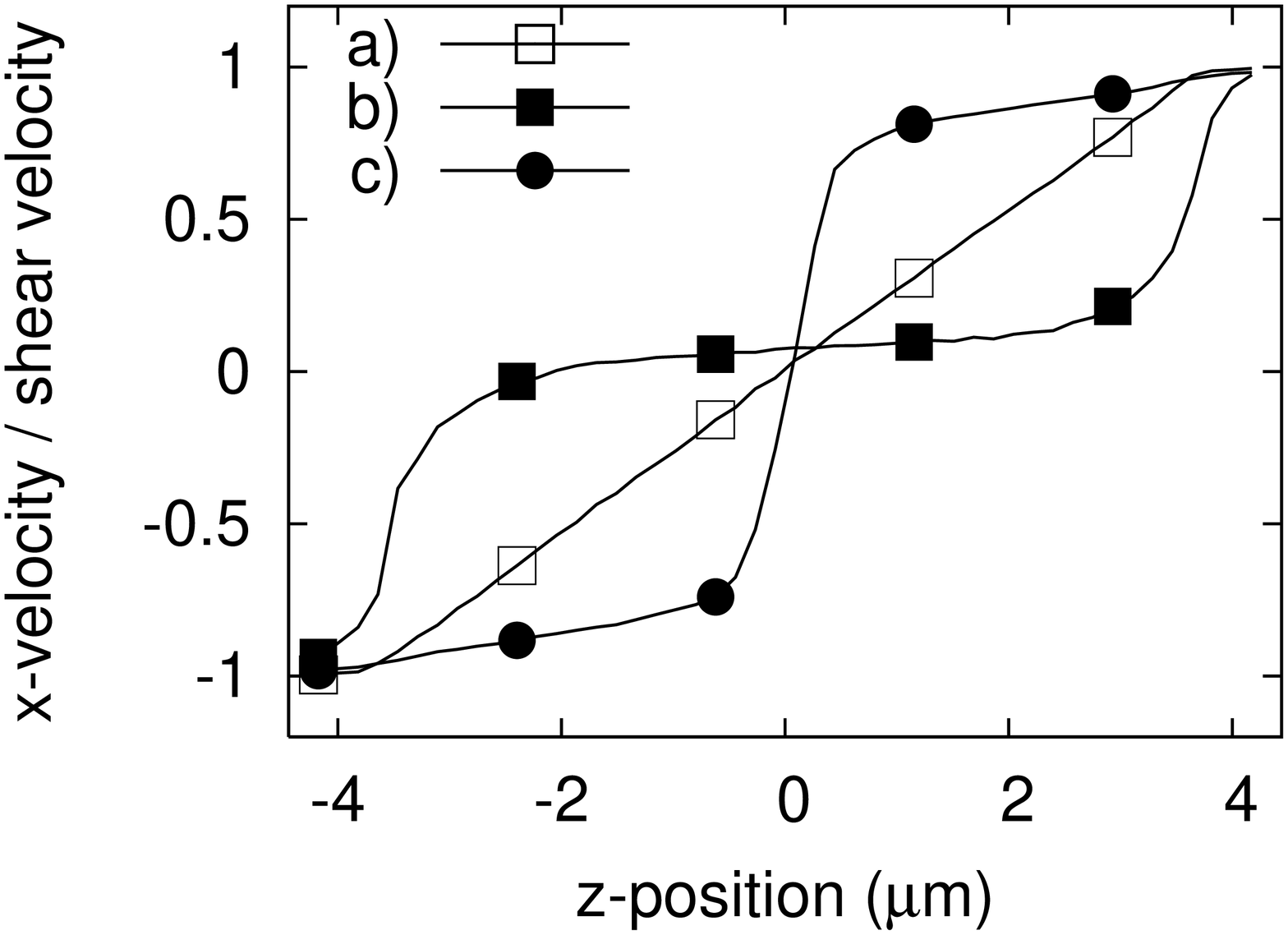}
\end{center}
\caption{Profiles of tangential velocity component in normal direction: 
a) Linear profile in the suspended regime, state $A$ of Fig.\ \ref{A7:figm_phasediag} ($I=3\,$mmol/l) at  $\dot\gamma=500/$s) \newline
b) Cluster formation in state $C$ ($I=25\,$mmol/l) at $\dot\gamma=100/$s. 
In principle one could determine the viscosity of one single cluster from the central plateau, 
but this is not the viscosity found in experiments. There, one measures the viscosity of a paste
consisting of many of these clusters \newline
c) Same as case b) but with higher shear rate (500/s). Hydrodynamic forces are large enough to break the cluster into two pieces. 
The velocity axis is scaled with the shear velocity $v_S$ for better comparability}
\label{A7:figm_VxProfiles}
\end{figure}

\begin{figure*}
\begin{center}
\mbox{{\bf a)}\includegraphics[scale=0.15]{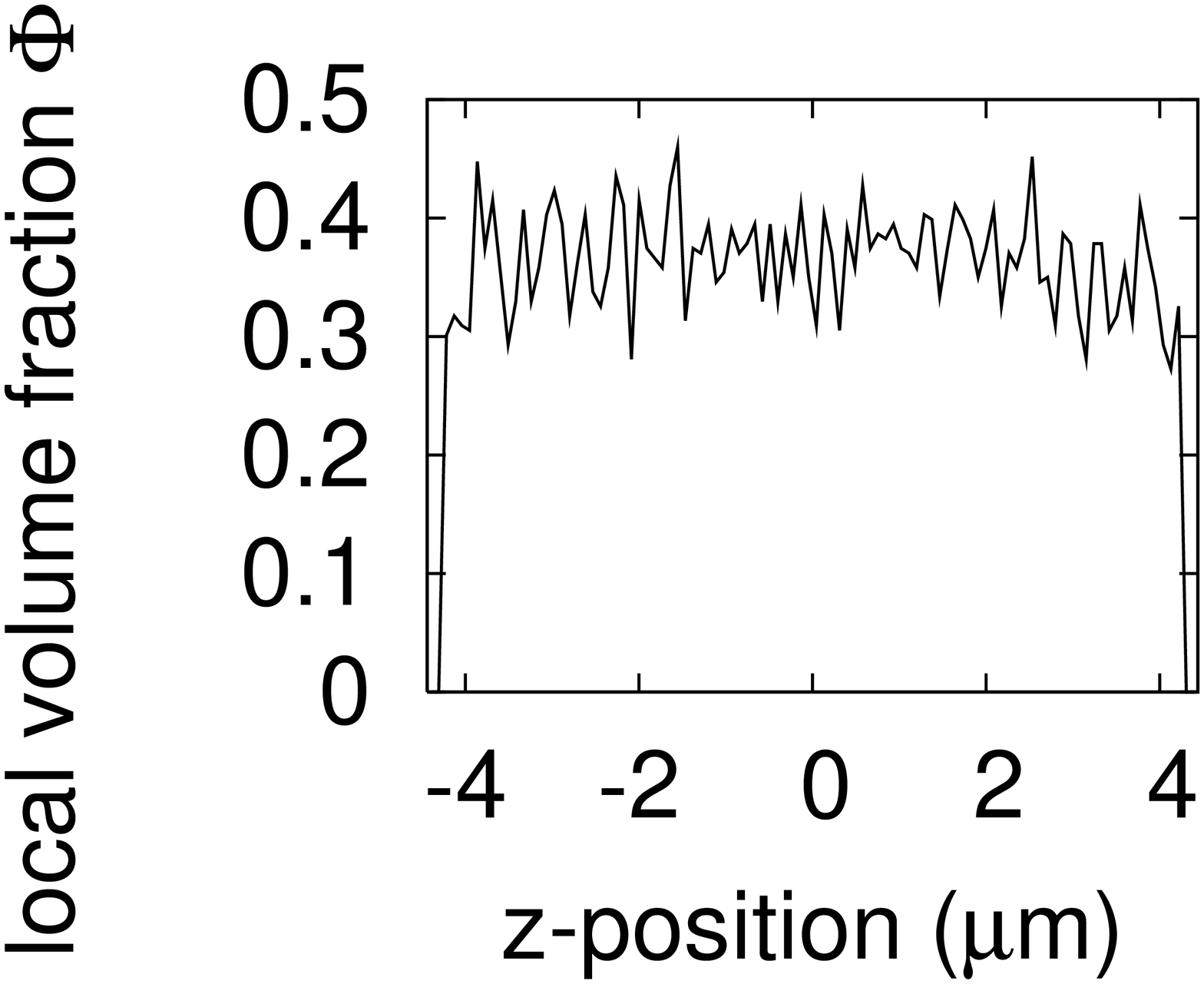}
{\bf b)}\includegraphics[scale=0.15]{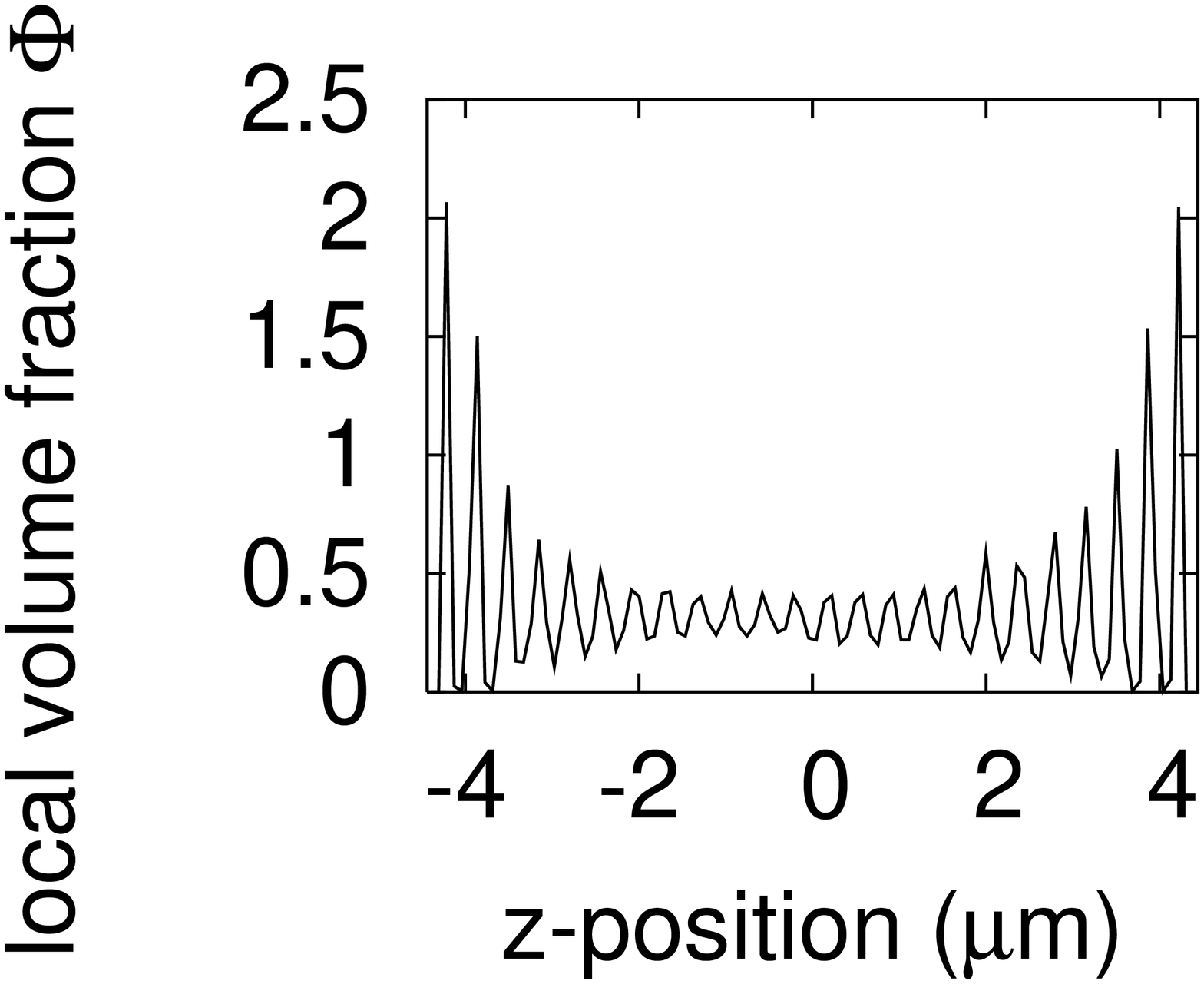}
{\bf c)}\includegraphics[scale=0.15]{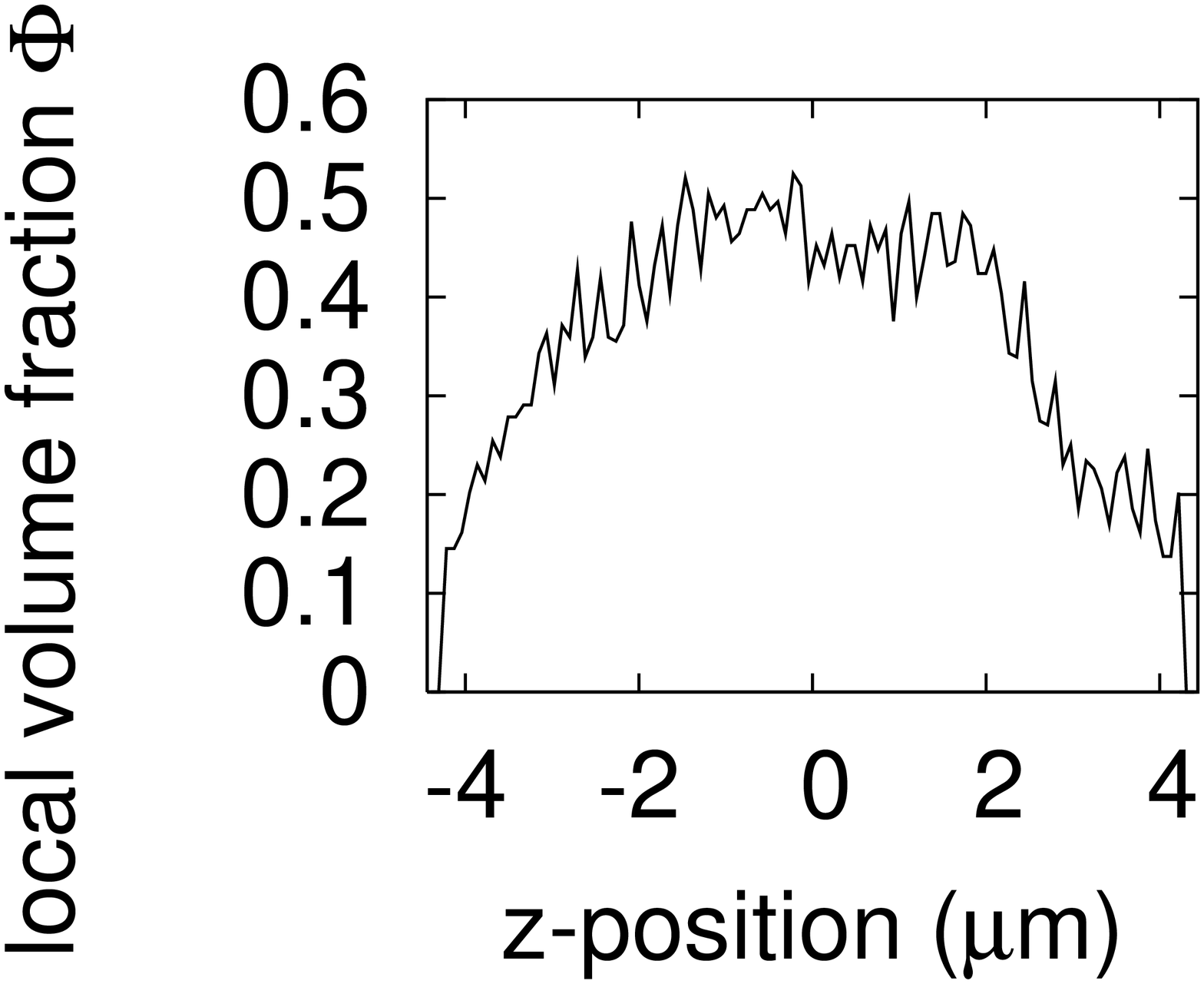}
}
\end{center}

\caption{Density profiles: \textbf{a)} Suspended case: State $A$ in Fig.\ \ref{A7:figm_phasediag} ($I=3\,$mmol/l), at low shear rates ($\dot\gamma=50/$s). The density distribution is homogeneous \newline
\textbf{b)} Shear induced layer formation: This is state $A$ as in graph a) of this figure, but for a high shear rate ($\dot\gamma=1000/$s) \newline
\textbf{c)} Strong attractive forces in state $C$ ($I=25\,$mmol/l): For low shear rates ($\dot\gamma=50/$s) only one central cluster is formed, which is deformed slowly
}
\label{A7:figm_densityProfiles}
\end{figure*}

\begin{figure}
\begin{center}
\includegraphics[scale=0.3]{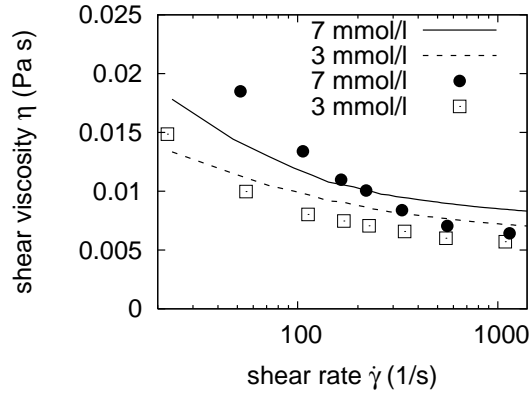}
\end{center}
\caption{Comparison between simulation and experiment: viscosity in dependence of the shear rate
for the states $A$ ($I=3\,$mmol/l) and $B$ ($I=7\,$mmol/l) of Fig.\ \ref{A7:figm_phasediag}. 
Note: shear thinning is more pronounced for the slightly attractive 
interactions in state $B$ than for the suspended state $A$. Lines denote experimental data 
\cite{A7:Reinshagen06}, points are results from our simulations}
\label{A7:figm_viscosity}
\end{figure}

In each of the three phases a typical velocity profile of the shear flow
occurs.  In the suspended phase one finds a linear velocity profile
(Fig.\ \ref{A7:figm_VxProfiles}a)) with nearly Newtonian flow. The
particles are distributed homogeneously, thus the density profile is
structureless (Fig.\ \ref{A7:figm_densityProfiles}a)).  The motion of the
particles is only weakly coupled by the hydrodynamic forces.  At high
enough shear rates ($\dot\gamma > 500$) the  particles arrange in layers
parallel to the shear plane, as can be seen in the density profile Fig.\
\ref{A7:figm_densityProfiles}b), too.  This arrangement minimizes
collisions between the particles. As a result, the shear viscosity
descents as shown in Fig.\ \ref{A7:figm_viscosity}, which we discuss
in more detail below.  Shear induced layer formation has been reported in
literature for different systems. Vermant and Solomon have reviewed this
topic recently~\cite{A7:Vermant05}.

\label{A7:secm_viscosity}
In the clustered phase, the clusters move in the fluid as a whole. They are deformed, but
since the inter-particle forces are stronger than the hydrodynamic forces, the 
cluster moves more like a solid body than like a fluid. 
Often there is one big cluster that spans the whole system. The density profile 
(Fig.\ \ref{A7:figm_densityProfiles}c)) increases in the central region and decays at the 
regions close to the border, since particles from there join the central cluster.
When averaging the velocity profile in the shear flow, one finds a very small velocity 
gradient in the center of the shear cell and fast moving particles close to the wall, 
where the shear is imposed (Fig.\ \ref{A7:figm_VxProfiles}b)). The velocity profile is 
non-linear on the length scale of the simulations. 
In the experiment the physical dimensions are much larger and therefore the velocity profile
can become approximately linear again if the system consists of many large clusters.
However, due to the computational effort in simulations 
it is today impossible to measure the shear viscosity for these strongly inhomogeneous systems.

Closer to the phase border clusters can then be broken up into small pieces by the 
hydrodynamic forces at least for high shear rates. In state $C$ of Fig.\ \ref{A7:figm_phasediag}
this happens for the first time at $\dot\gamma = 500/$s, so that one can find two clusters in 
the system moving in opposite directions. The velocity profile of this case is shown in
Fig.\ \ref{A7:figm_VxProfiles}c). For even higher shear rates or closer to the phase 
border (e.g. state $B$), the clusters are broken into smaller pieces. Then, they 
move in the shear flow with an approximately linear velocity profile. 
Due to van der Waals attraction the system resists with stronger shear forces and the 
viscosity is higher than in the suspended case (Fig.\ \ref{A7:figm_viscosity}). 

In Fig.\ \ref{A7:figm_viscosity} the simulation results are shown together with the experimental
results, both for the two cases of a slightly clustered system in state $B$ ($I = 7\,$mmol/l) 
and a suspension (state $A$, $I = 3\,$mmol/l). 
For the suspension (state $A$) the viscosity decreases with the shear 
rate (``shear thinning''). The experimental data and the simulation are 
consistent within the accuracy of our model. 
There are several 
reasons for which our model does not fit exactly the measurements:
Even though we use a charge regulation model to determine the input parameters 
for the DLVO potentials, microscopic properties like the surface density 
of sites, where ions can be adsorbed on the surface of the colloidal particle, 
have to be determined indirectly. Measurements of the $\zeta$-potential 
in certain conditions provide data to fit the unknown microscopic parameters.
Furthermore, we have monodisperse spheres, which is another simplification in our model. 

For the slightly clustered case (state $B$) an increase of the shear viscosity, 
compared to the suspended case, can be observed in the experiment as well as in the simulations. 
Shear thinning becomes more pronounced, because clusters are broken up,
as mentioned above. However, the shear rate dependence is stronger in the simulations
than in the experiment. This can be the first indication of finite size effects. 
We have studied the dependence of the simulated shear viscosity in dependence of the system 
size. The effect is most important for low shear rates.

\section{Transport Phenomena and Structuring in Suspensions: Lattice-Boltzmann Simulations}
For industrial applications, systems with rigid boundaries, e.g. a pipe
wall, are of particular interest since structuring effects might occur in
the solid fraction of the suspension. Such effects are known from dry
granular media resting on a plane surface or gliding down an
inclined chute \cite{A7:mijatovic02a,A7:poeschel93a}. In addition, the wall
causes a demixing of the solid and fluid components which might have an
unwanted influence on the properties of the suspension. Near the wall one
finds a thin lubrication layer which contains almost no particles and
causes a so-called ``pseudo wall slip''. Due to this slip the suspension
can be transported substantially faster and less energy is dissipated. 

We expect structuring close to a rigid wall at much smaller concentrations
than in granular media because of long-range hydrodynamic interactions.
In \cite{A7:jens-komnik-herrmann:2004}, we study these effects by the means of particle
volume concentrations versus distance to the wall.

\subsection{The Lattice-Boltzmann Method\label{A7:sec:simulation_method}}
The lattice-Boltzmann method is a simple scheme for simulating the dynamics of
fluids. By incorporating solid particles into the model fluid and
imposing the correct boundary condition at the solid/fluid interface,
colloidal suspensions can be studied. Pioneering work on the development
of this method has been done by Ladd et al.
\cite{A7:Ladd94, A7:Ladd94b,A7:Ladd01} and we use their approach to model
sheared suspensions near solid walls.

The lattice-Boltzmann (hereafter LB) simulation technique which is
based on the well-established connection between the dynamics of a dilute
gas and the Navier-Stokes equations \cite{A7:chapman60a}.
We consider the time evolution of the one-particle velocity distribution
function $n(\vec{r},\vec{v},t)$, which defines the density of particles
with velocity $\vec{v}$ around the space-time point $(\vec{r},t)$. By
introducing the assumption of molecular chaos, i.e. that successive binary
collisions in a dilute gas are uncorrelated, Boltzmann was able to derive
the integro-differential equation for $n$ named after him \cite{A7:chapman60a}
\begin{equation}
\partial_tn+\vec{v}\cdot\nabla n=\left(\frac{dn}{dt}\right)_{\mathrm coll}\;,
\end{equation}
where the left hand side describes the change in $n$ due to collisions.

The LB technique arose from the
realization that only a small set of discrete velocities is necessary to
simulate the Navier-Stokes equations \cite{A7:frisch86a}. Much of the
kinetic theory of dilute gases can be rewritten in a discretized
version. The time evolution of the distribution functions $n$ is described by a
discrete analogue of the Boltzmann equation \cite{A7:Ladd01}: 
\begin{equation}\label{A7:eq:diskr-boltzmann}
n_i(\vec{r}+\vec{c}_i\Delta t, t+\Delta t) = n_i(\vec{r}, t) +
\Delta_i(\vec{r},t)\;,
\end{equation}
where $\Delta_i$ is a multi-particle collision term.
Here, $n_i(\vec r,t)$ gives the density of particles with velocity
$\vec{c}_i$ at $(\vec r,t)$.  In our simulations, we use 19
different discrete velocities $\vec{c}_i$.
The hydrodynamic fields, mass density $\varrho$, momentum density
$\vec{j}=\varrho\vec{u}$, and momentum flux ${\Pi}$, are moments of this
velocity distribution:
\begin{align}
\varrho&=\sum_in_i\;,&\vec{j}&=\varrho\vec{u}=\sum_in_i\vec{c}_i\;,&{\Pi}&=%
\sum_in_i\vec{c}_i\vec{c}_i\;.
\end{align}
We use a linear collision operator,
\begin{equation}
\Delta_i(r,t)=M_{ij}(n_j - n_j^{eq})\;,
\end{equation}
where $M_{ij}\equiv\frac{\partial\Delta_i(n^{eq})}{\partial n_j}$ is the
collision matrix and $n_i^{eq}$ the equilibrium distribution \cite{A7:chen98a},
which determines the scattering rate between directions $i$ and $j$.
For mass and momentum conservation, $M_{ij}$ satisfies the constraints
\begin{align}
\sum_{i=1}^MM_{ij} &= 0\;,&\sum_{i=1}^M\vec{e}_iM_{ij} = 0\;.
\end{align}
We further assume that the local particle distribution relaxes to an
equilibrium state at a single rate $\tau$ and obtain the lattice BGK
collision term \cite{A7:bhatnagar54a}
\begin{equation}\label{A7:eq:normalcollision}
\Delta_i=-\frac{1}{\tau}(n_i - n_i^{eq})\;.
\end{equation}
By employing the Chapman-Enskog expansion \cite{A7:chapman60a,A7:frisch87a}
it can be shown that the equilibrium distribution
\begin{equation}
n_i^{eq}=\varrho\omega^{c_i}\left[1+3\vec{c}_i\cdot\vec{u}+\frac{9}{2}(\vec{c}_i
\cdot \vec{u})^2-\frac{3}{2}u^2\right]\;,
\end{equation}
with the coefficients of the three velocities
\begin{align}\label{A7:eq:omegas}
\omega^0 &= \frac{1}{3}\;,&\omega^1 &= \frac{1}{18}\;,&\omega^{\sqrt{2}} &=
\frac{1}{36}\;,
\end{align}
and the kinematic viscosity \cite{A7:Ladd01}
\begin{equation}\label{A7:eq:viskositaet}
\nu = \frac{\eta}{\varrho_f} = \frac{2\tau - 1}{9}\;,
\end{equation}
properly recovers the Navier-Stokes equations
\begin{equation}
\frac{\partial u}{\partial t} + (u\nabla)u = -\frac{1}{\varrho}\nabla p +
\frac{\eta}{\varrho}\Delta u\;,\qquad\nabla u = 0\;.
\end{equation}

\subsection{Fluid-Particle Interactions}
To simulate the hydrodynamic interactions between solid particles in
suspensions, the lattice-Boltzmann model has to be modified to incorporate the
boundary conditions imposed on the fluid by the solid particles.
Stationary solid objects are introduced into the model by replacing the 
usual collision rules (Equation \eqref{A7:eq:normalcollision}) at a specified set
of boundary nodes by the ``link-bounce-back'' collision rule \cite{A7:nguyen02}.
When placed on the lattice, the boundary surface cuts some of
the links between lattice nodes. The fluid particles moving along these links
interact with the solid surface at boundary nodes placed halfway along the
links. Thus, a discrete representation of the surface is obtained, which
becomes more and more precise as the surface curvature gets smaller and 
which is exact for surfaces parallel to lattice planes.

Since the velocities in the lattice-Boltzmann model are discrete, boundary
conditions for moving suspended particles cannot be implemented directly.
Instead, we can modify the density of returning particles in a way that
the momentum transferred to the solid is the same as in the continuous
velocity case.
This is implemented by introducing
an additional term $\Delta_b$ in \eqref{A7:eq:diskr-boltzmann}
\cite{A7:Ladd94}:
\begin{equation}\label{A7:eq:moving-collision-rule}
\Delta_{b,i}=\frac{2\omega^{c_i}\varrho_i\vec{u}_i\cdot\vec{c}_i}{c_s^2}\;,
\end{equation}
with $c_s$ being the velocity of sound and
coefficients $\omega^{c_i}$ from \eqref{A7:eq:omegas}.

To avoid redistributing fluid mass from lattice nodes being covered or
uncovered by solids, we allow interior fluid within closed surfaces. Its
movement relaxes to the movement of the solid body on much shorter time
scales than the characteristic hydrodynamic interaction \cite{A7:Ladd94}.

If two particle surfaces approach each other within one lattice spacing,
no fluid nodes are available between the solid surfaces.
In this case, mass is not conserved anymore since boundary updates
at each link produce a mass transfer $\Delta_ba^3$ ($a\equiv$cell size)
across the solid-fluid interface \cite{A7:Ladd94}.
The total mass transfer for any closed surface is zero,
but if some links are cut by two surfaces, no solid-fluid
interface is available anymore.
Instead, the surface of each particle is not closed at the solid-solid
contacts anymore and mass can be transferred in-between suspended particles.
Since fluid is constantly added or removed from the individual particles,
they never reach a steady state.
In such cases, the usual boundary-node update procedure
is not sufficient and a symmetrical procedure which takes account of both
particles simultaneously has to be used \cite{A7:Ladd94b}.
Thus, the boundary-node velocity is taken to be
the average of that computed from the velocities of each particle.
Using this velocity, the fluid populations are updated
(Equation \eqref{A7:eq:moving-collision-rule}), and the force is
computed; this force is then divided equally between the two particles.

\index{lubrication interactions}
If two particles are in near contact, the fluid flow in the gap cannot be
resolved by LB. For particle sizes used in our simulations ($R < 5a$), the
lubrication breakdown in the calculation of the hydrodynamic interaction
occurs at gaps less than $0.1R$ \cite{A7:nguyen02}. This effect 
``pushes'' particles into each other.

To avoid
this force, which should only occur on intermolecular distances,
we use a lubrication correction method described
in \cite{A7:nguyen02}. For each pair of particles a force
\begin{align}
\vec{F}_\text{lub}&=
-6\pi\eta\frac{R_1R_2}{(R_1+R_2)^2}\left(\frac{1}{h}-\frac{1}{h_N}
\right)\vec{u}_{12}\cdot\frac{\vec{r}_{12}}{\lvert\vec{r}_{12}\rvert}\;,&h <
h_N
\end{align}
is calculated, where
$\vec{u}_{12}=\vec{u}_1-\vec{u}_2, h=\lvert\vec{r}_{12}\rvert-R_1-R_2$
is the gap between the two surfaces and a cut off distance $h_N =
\frac{2}{3}a$ \cite{A7:Ladd01}. For particle-wall contacts we apply the same
formula with $R_2\rightarrow\infty$ and $h=\lvert\vec{r}_{12}\rvert-R_1$.
The tangential lubrication can also be taken into account, but since it has
a weaker logarithmic divergence and its breakdown does not lead to serious
problems, we do not include it in our simulations.

\subsection{Particle Motion}
The particle position and velocity are calculated using Newton's equations
in a similar manner as in section on SRD simulations. To avoid repetition,
the reader is referred to Sect.\ \ref{A7:SRD-MD}. However, particles do
not feel electrostatic interactions, but behave like hard spheres in the
case presented in this section.

\subsection{Simulations\label{A7:LB:sec:simulations}}
The purpose of our simulations is the reproduction of rheological
experiments on computers. We simulate a representative volume
element of the experimental setup and compare our calculations
with experimentally accessible data, i.e. density profiles, time
dependence of shear stress and shear rate. We also get experimentally
inaccessible data from our simulations like translational and rotational
velocity distributions, particle-particle and particle-wall interaction
frequencies. The experimental setup consists of a rheoscope with two
spherical plates, which distance can be varied. The upper plate can be
rotated either by exertion of a constant force or with a constant
velocity, while the complementary value is measured simultaneously.  The
material between the rheoscope plates consist of glass spheres suspended
in a sugar-water solution. The radius of the spheres varies between $75$
and $150$ $\umu$m.  For our simulations we assume an average particle
radius of 112.5 $\umu$m. The density and viscosity of the sugar solution
can also be changed. We simulate only the behavior of a representative
volume element which has the experimental separation between walls, but a
much lower extension in the other two dimensions than the experiment.  In
these directions we employ periodic boundary conditions for particles and
for the fluid.

Shearing is implemented using the ``link-bounce-back'' rule with an
additional term $\Delta_{b,i}$ at the wall in the same way as already
described for particles
(Equation \eqref{A7:eq:moving-collision-rule} with $\vec{u}_i$ now being
the velocity of the wall).

To compare the numerical and experimental results, we need to find
characteristic dimensionless quantities of the experiment which then
determine the simulation parameters. For this purpose we use the ratio of
the rheoscope height and the particle size $\lambda$, the particle
Reynolds number $\Re$ and the volume fraction of the particles $\phi$. The
simulation results are provided with units by calculating the length of
the lattice constant $a$ and the duration of one time step as described in
\cite{A7:jens-komnik-herrmann:2004}.

\subsection{Results\label{A7:lb:sec:results}}
\begin{figure}
\centering
{\fboxsep=0pt
\fbox{\includegraphics[width=40mm]{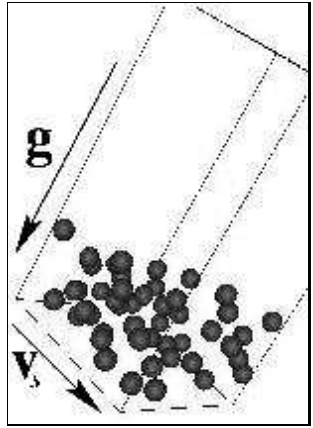}}
}
\caption{\label{A7:fig:50kugeln}A snapshot of a suspension with 50 spheres
(radius $R = 1.125\cdot10^{-4}\text{ m}$, mass $m=7.7\cdot10^{-8}\text{ kg}$) 
at time $t = 729\text{ s}$.
The volume of the simulated system is
$1.83\cdot10^{-3}\times1.83\cdot10^{-3}\times3.375\cdot10^{-3}\text{ m}=
11.3025\cdot10^{-9}\text{ m}^3$,
acceleration of gravity $g = 0.80\text{ m/s}^2$, and shear velocity
$v_s = 3.375\cdot10^{-2}\text{ m/s}$.
The fluid has a viscosity
$\eta=450\text{ mPa}\cdot\text{s}$ and density
$\varrho_f=1446\frac{\text{kg}}{\text{m}^3}$.
This visualization is a typical example for a system that has reached a
steady state: All particles have fallen to the ground due to the exerted
gravitational force and most of the system has no particles
\cite{A7:jens-komnik-herrmann:2004}}
\end{figure}

Figure \ref{A7:fig:50kugeln} shows a snapshot of a suspension with 50 spheres
after $5772500$ time steps which are equivalent to $729\text{ s}$.
The vector $\vec{g}$ represents the direction of gravity and $\vec{v}_{S}$
depicts the velocity of the sheared wall.

The particles feel a gravitational acceleration $g=0.8\text{ m/s}^2$,
have a mass $m=7.7\cdot10^{-8}\text{ kg}$,
a Reynolds number $\Re=4.066875\cdot10^{-4}$,
and a radius $R=1.125\cdot10^{-4}\text{ m}$.
The system size is
$1.83\cdot10^{-3}\times1.83\cdot10^{-3}\times3.375\cdot10^{-3}\text{ m}$
which corresponds to a lattice size of $32\times32\times59$.
The density of the fluid is set to $\varrho_f=1446\frac{\text{kg}}{\text{m}^3}$
and its viscosity is $\eta=450\text{ mPa}\cdot\text{s}$.
The walls at the top and the bottom are sheared with a relative velocity
$v_s = 3.375\cdot10^{-2}\text{ m/s}$.
Figure \ref{A7:fig:50kugeln} is a
representative visualization of our simulation data and demonstrates that
after the system has reached its steady state, all particles have fallen
to the ground due to the influence of the gravitational force.
Most of the simulation volume is free of particles.

\begin{figure*}
\includegraphics[width=60mm]{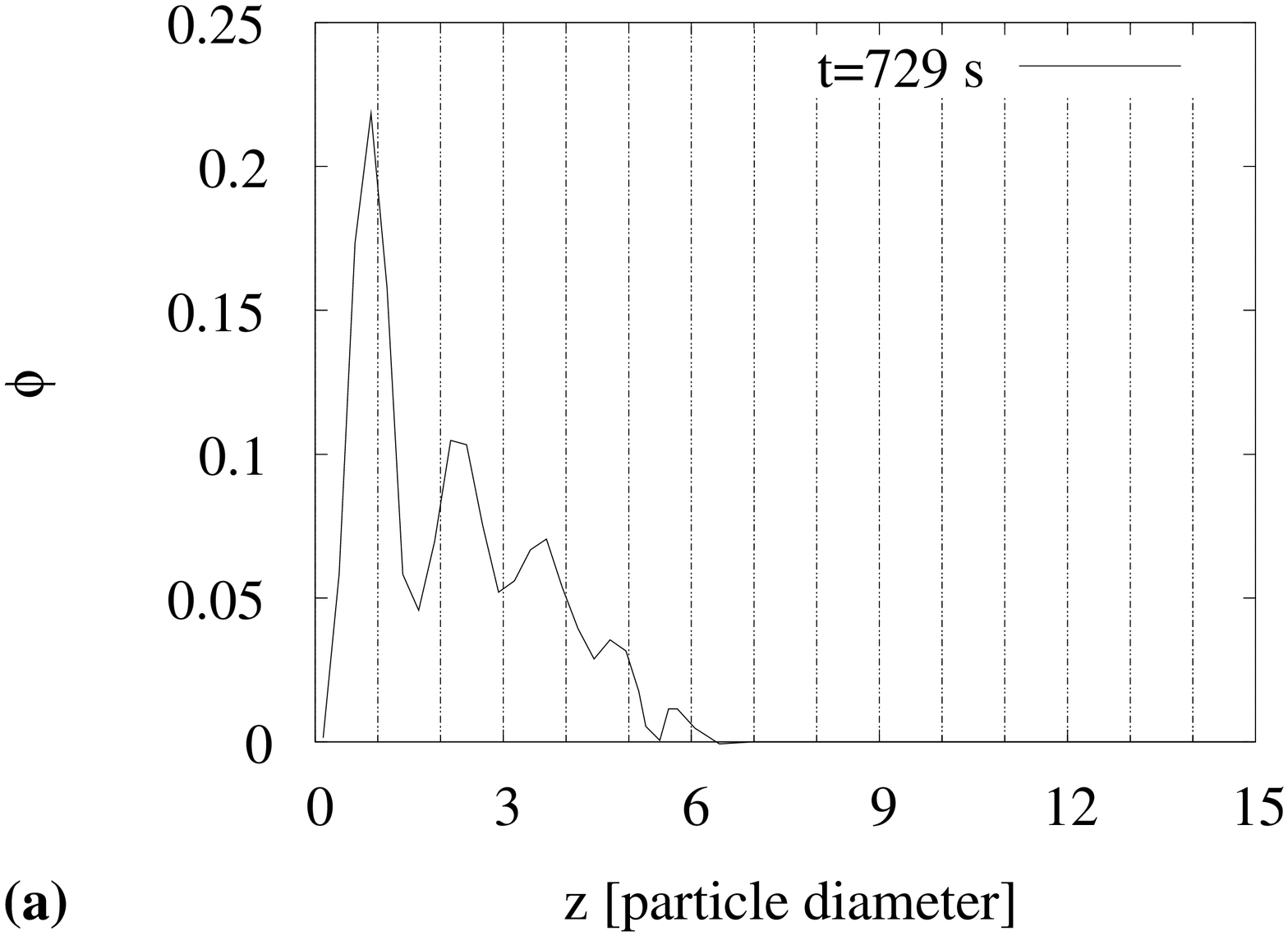}
\hfill
\includegraphics[width=60mm]{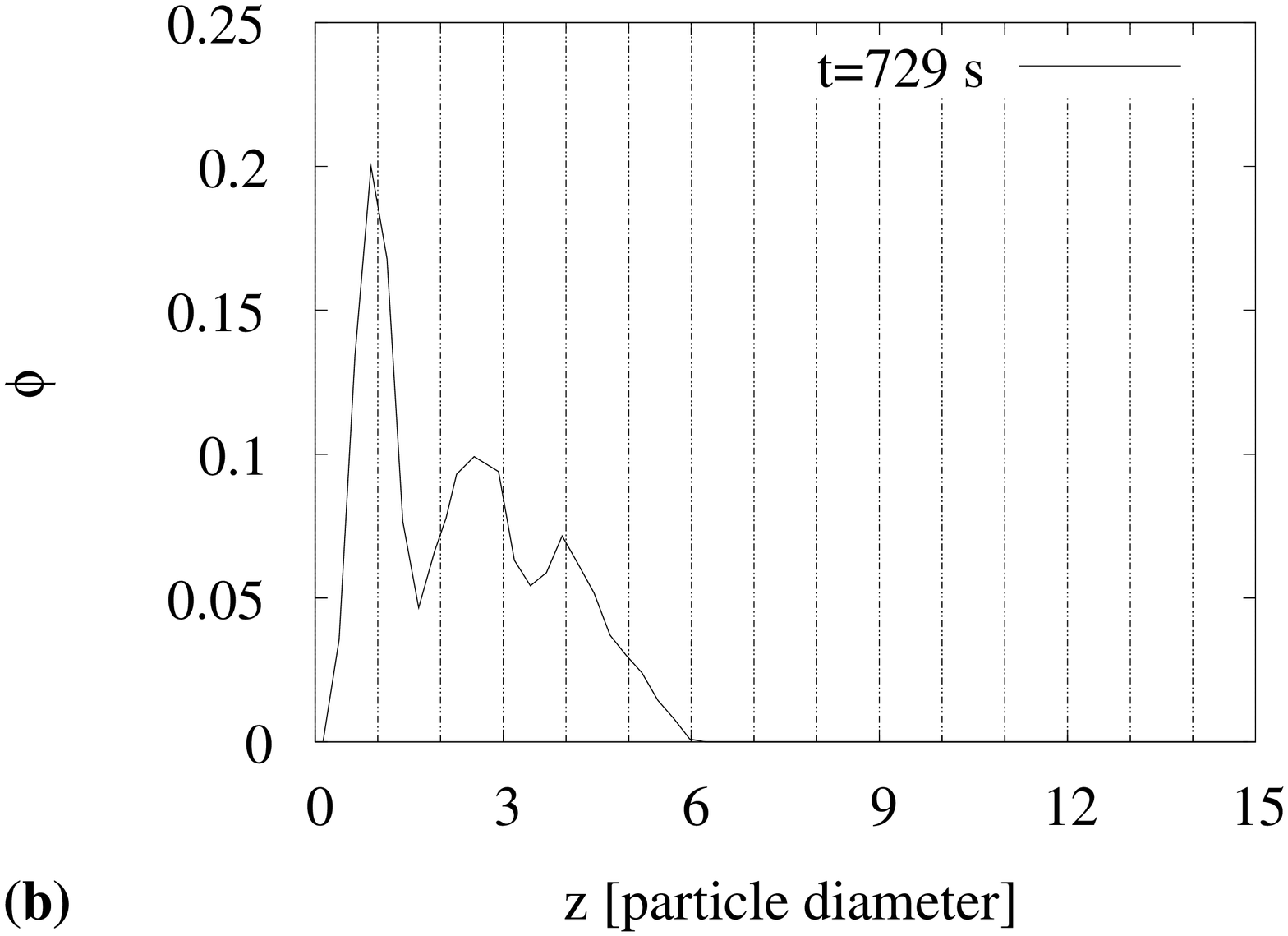}
\caption{\label{A7:fig:50dichten}
Density profiles from simulations with two different shear rates
$\gamma=10\text{ s}^{-1}$ (\textbf{a}) and $\gamma=1\text{ s}^{-1}$
(\textbf{b}).
Other parameters are equal to those given in Fig.\ \ref{A7:fig:50kugeln}.
(\textbf{a}) shows five peaks with separations about one particle diameter,
which reveal the forming of particle layers.  The number of particles per
layer is decreasing with increasing distance to the wall, and the change
in particle numbers is caused by gravity which is directed perpendicular
to the wall at $z=0$.  Although we used the same gravity and particle
numbers, there are only three peaks in (\textbf{b}) and their width is
higher than in (\textbf{a}), demonstrating that the structuring effects
strongly relate to the shear rate
}
\end{figure*}
In order to quantitatively characterize structuring effects, we calculate
the particle density profile of the system by dividing the whole system
into layers parallel to the walls and calculating a partial volume $V_{ij}$ for
each particle $i$ crossing such a layer $j$. The scalar $V_{ij}$ is given
by the volume fraction of particle $i$ that is part of layer $j$:
\begin{equation}
V_{ij}=
\pi\left(R^2\left(R_{ij}^\text{max}-R_{ij}^\text{min}\right)-
\frac{1}{3}\left(R_{ij}^\text{max}-R_{ij}^\text{min}\right)\right)
\end{equation}
If the component $r_{i,z}$ perpendicular to the wall of the radius vector
$\vec{r}_i$ of the center of sphere $i$ lies between
$r_j^\text{min}$ and $r_j^\text{max}$, we have
\begin{align}
\notag r_j^\text{min}&=\left(j-\frac{1}{2}\right)\Delta L_z-R\;,\\
\notag r_j^\text{max}&=\left(j+\frac{1}{2}\right)\Delta L_z+R\;,
\end{align}
and
\begin{align}
\notag R_{ij}^\text{max}&=\left\{\begin{array}{ll}
R&\text{if }r_{i,z}+R<r_j^\text{max}\\
r_j^\text{max}-r_{i,z}&\text{else}\\
\end{array}\right.\;,\\
\notag R_{ij}^\text{min}&=\left\{\begin{array}{ll}
-R&\text{if }r_{i,z}-R>r_j^\text{min}\\
r_j^\text{min}-r_{i,z}&\text{else}\\
\end{array}\right.\;.
\end{align}
Finally, the sum of all weights associated with a layer is divided by the
volume of the layer
\begin{align}
\phi_j&=\frac{1}{L_x\cdot L_y\cdot\Delta L_z}\sum\limits_{i=1}^Nv_{ij},&
\Delta L_z&=\frac{L_z}{M}\;,
\end{align}
with $L_x, L_y$ being the system dimensions between periodic boundaries,
$L_z$ the distance between walls, $M$ the number of layers,
and $\Delta L_z$ the width of a single layer.

Density profiles calculated by this means for systems with two different
shear rates $\gamma=10\text{ s}^{-1}$ and $\gamma=1\text{ s}^{-1}$
are presented in Fig.\ \ref{A7:fig:50dichten}.
All other parameters are equal to the set given in
the last paragraph. The peaks in Fig.\ \ref{A7:fig:50dichten} demonstrate
that at certain distances from the wall the number of particles is
substantially higher than at other positions.
The first peak in both figures is slightly below
one particle diameter, which can be explained by a lubricating fluid
film between the first layer and the wall which is slightly thinner than
one particle radius. Due to the small amount of particles, time dependent
fluctuations of the width of the lubricating layer cannot be neglected and
a calculation of the exact value is not possible.
The five peaks in Fig.\ \ref{A7:fig:50dichten}\textbf{a} have similar
distances which are equal to one particle diameter.
These peaks can be explained by closely packed parallel layers of
particles. Due to the linear velocity profile in $z$-direction of the
fluid flow, every layer adopts the local velocity of the fluid resulting
in a relative velocity difference between two layers of about $2R\gamma$.
These layers stay stable in time with
only a small number of particles being able to be exchanged between them. 

Figure \ref{A7:fig:50dichten}\textbf{b} only shows three peaks with larger
distances than in Fig.\ \ref{A7:fig:50dichten}\textbf{a}. However, the
average slope of the profile is identical for both shear rates.
For smaller shear rates, velocity differences between individual layers
are smaller, too. As a result, particles feel less resistance while moving
from one layer to another. Every inter-layer transition distorts the well
defined peak structure of the density distribution resulting in only three
clearly visible peaks in Fig.\ \ref{A7:fig:50dichten}\textbf{b}.

With changing time, the first peak stays constant for both shear rates.
The shape, number and position of all other peaks is slightly changing
in time.

\begin{figure}
\centering
{{\includegraphics[width=60mm]{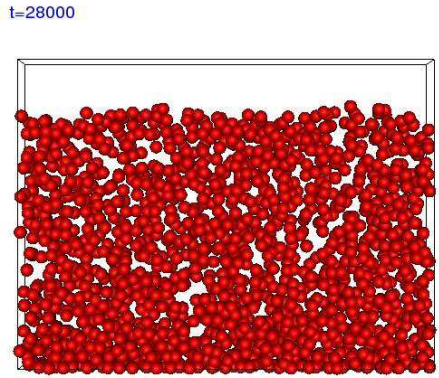}}
}
\caption{\label{A7:fig:1536kugeln}A snapshot of a suspension with 1536
spheres after 28000 timesteps used to gain statistics of particle velocity distributions}
\end{figure}
%
We are currently investigating the occurrence of 
non-Gaussian velocity distributions of particles for higher  
particle densities and higher shear rates. For this, improvements of
the method are mandatory in order to prevent instabilities of the
simulation. By utilizing an implicit scheme for the update of the particle
velocities \cite{A7:Ladd01,A7:nguyen02} we are able to overcome artefacts
caused by numerical inaccuracies at high volume fractions or shear rates.
Figure \ref{A7:fig:1536kugeln} shows a snapshot of a system containing
1536 particles after 28000 timesteps.

The lattice Boltzmann has been extended in order to include thermal
fluctuations \cite{A7:Ladd01,A7:Cates05}. With these modifications the
method is another candidate to simulate suspensions where Brownian motion
cannot be neglected.

\section{Plug Conveying in Vertical or Horizontal Tubes: a Coarse Grained
Model for the Fluid Flow}
\newcommand{\pd}[2]{\frac{\partial {#1}}{\partial {#2}}}
\newcommand{\grad}{\vec{\nabla}}

\subsection{Model Description}
\index{pneumatic transport}
\index{plug conveying}
Another approach to modeling two phase flow is to course-grain the
fluid, so that it is resolved on a length scale larger than the
grains.  The advantage is that much larger systems can be treated, but the
disadvantage is that this coarse-graining is justified only in certain
situations.  One of those situations is when the density of the fluid is
small compared to that of the grains, and the Reynolds number of the grains
is small.  It is then possible to neglect the inertia of the fluid, which means
that all momentum is contained in the grains. The fluid transfers momentum 
between grains, but stores no momentum itself.  And when the Reynolds
number of the grains is small, one can treat the granulate phase as a 
moving porous medium.  In the following, we present the model in more
detail.

\subsubsection{Gas Model}
The model for the gas simulation was first introduced by
McNamara and Flekk{\o}y \cite{A7:MCN0003} and has been implemented for
the two-dimensional case
to simulate the rising of bubbles within a fluidized bed.
We developed a three-dimensional version of this algorithm.

The algorithm is based on the mass conservation of the gas
and the granular medium.
Conservation of grains implies that
the density $\varrho_p$ of the granular medium obeys
\begin{equation}\label{equ:rhop}
\pd{\varrho_p}{t}+\vec{\nabla}\cdot(\vec{u}\varrho_p)=0\;,\qquad
\varrho_p=\varrho_{s}(1-\phi)\;,
\end{equation}
where $\varrho_p$ is the mass density of the material making up the particles,
the porosity of the medium is $\phi$
(i.e. the fraction of the space available to the gas),
and the velocity of the granulate is $\vec{u}$.

The mass conservation equation for the gas is
\begin{equation}\label{equ:rhog}
\pd{\varrho_g}{t}+\vec{\nabla}\cdot(\vec{v}_g\varrho_g)=0\;,
\qquad \varrho_g\propto \phi P\;,\quad\qquad 
\end{equation}
where $\varrho_g$ is the mass density of the gas
averaged over the total volume of the granular medium
 and $\vec{v}_g$ its velocity.
This equation
can be transformed into a differential equation for the gas pressure $P$
using the ideal gas equation, together with the assumption of uniform
temperature.

The velocity $\vec{v}_g$ of the gas is related to
the granulate velocity $\vec{u}$ through the d'Arcy relation:
\begin{equation}\label{eqn:darcy}
-\vec{\nabla}P=\frac{\eta}{\kappa(\phi)}\phi(\vec{v}_g-\vec{u})\;,
\end{equation}
where $\eta$ is the dynamic viscosity of the air and $\kappa$ is the
permeability of the granular medium. This relation was first given
by d'Arcy in 1856 \cite{A7:DAR0094}. The d'Arcy relation is preferred here 
over the Ergun equation, because it is linear in the velocity.
This makes the simplification steps done later possible.
For the permeability $\kappa$ the Carman-Kozeny relation \cite{A7:CAR0084}
was chosen,
which provides a relation between the porosity $\phi$, the particle diameter $d$
and the permeability of a granular medium of monodisperse spheres,
\begin{equation}
\kappa(\phi)=\frac{d^2\phi^3}{180(1-\phi)^2}\;.
\end{equation}
Combining \eqref{equ:rhop}, \eqref{equ:rhog} and \eqref{eqn:darcy}
results in a nonlinear differential equation for the gas pressure:
\begin{equation}
\phi(\pd{P}{t}+\vec{u}\vec{\nabla}P)=\vec{\nabla}(P\frac{\kappa(\phi)}{\eta}\vec{\nabla}P)-P\vec{\nabla}\vec{u}\;.
\end{equation}
After linearizing around
the normal atmospheric pressure $P_0$ the resulting differential equation
only depends on the relative pressure $P^\prime$ ($P=P_0+P^\prime$),
the porosity $\phi$ and the granular velocity $\vec{u}$,
which can be derived from the particle simulation,
and three constants: the viscosity~$\eta$, the particle diameter~$d$
and the pressure~$P_0$:
\begin{equation}
\pd{P^\prime}{t}=\frac{P_0}{\eta\phi}\vec{\nabla}(\kappa(\phi)\vec{\nabla}P^\prime)-\frac{P_0}{\phi}\vec{\nabla}\vec{u}\;.
\end{equation}
This differential equation can be interpreted as a diffusion equation with
a diffusion constant $D=\phi\kappa(\phi)/\eta$.
The equation is solved numerically, using a
Crank-Nickelson approach for the discretization. Each dimension is integrated
separately.

\subsubsection{Granulate Algorithm}
The model for the granular medium simulates each grain
individually using a discrete element simulation (DES).
For the implementation of the discrete element simulation we used 
a version of the molecular dynamics method described by 
Cundall~\cite{A7:CUN0172}.
The particles are approximated as monodisperse spheres, rotations in
three dimensions are taken into account.

The equation of motion for an individual particle is
\begin{equation}
m\ddot{\vec{x}}=m\vec{g}+\vec{F}_{c}-\frac{m\grad P}{\varrho_s(1-\phi)}\;,
\end{equation}
where $m$ is the mass of a particle, $\vec{g}$ the gravitation constant and
$\vec{F}_{c}$ the sum over all contact forces. The last term,
the drag force, is assumed to be a volume force
given by the pressure drop $\grad P$ and the local mass density of
the granular medium $\varrho_s(1-\phi)$, which is valid for monodisperse granular media.

The interaction between two particles in contact is given
by two force components:
a normal and a tangential component with respect to the particle surface.
The normal force is the sum of a repulsive elastic force (Hooke's law)
and a viscous damping.
The tangential force opposes the relative tangential motion and 
is proportional to the normal force (sliding Coulomb friction) or
proportional to the relative tangential velocity (viscous damping).
Viscous damping is used only for small relative tangential velocities.

\subsubsection{Gas-Grain Interaction}
The simulation method uses both a continuum and a
discrete element approach. While the gas algorithm
uses fields, which are discretized on a cubic grid,
the granulate algorithm describes particles in a continuum.
A mapping is needed for the algorithms to interact.
For the mapping a tent function $F(\vec{r})$ is used:
\begin{equation}
F(\vec{r})=f(x)f(y)f(z),\qquad f(x)=\begin{cases}
  1-|x/l|, &  |x/l| \le 1\;, \\
  0\;, &  1<|x/l|\;,
\end{cases}
\end{equation}
where $l$ is the grid constant used for the discretization
of the gas simulation.

For the gas algorithm the porosity $\phi_j$ 
and the granular velocity $\vec{u}_j$
must be derived from the particle positions $\vec{r}_i$ and velocities $\vec{v}_i$,
where $i$ is the index of particle and
$j$ is the index for the grid node.
The tent function distributes the particle properties around
the particle position smoothly on the grid:
\begin{equation}
\phi_j=1-\sum_i F(\vec{r}_i-\vec{r}_j)\;, \qquad
\vec{u}_j=\frac{1}{1-\phi_j}\sum_i \vec{v}_i F(\vec{r}_i-\vec{r}_j)\;,
\end{equation}
where $r_j$ is the position of the grid point 
and the sum is taken over all particles.

For the computation of the drag force on a particle
the pressure drop $\grad P_i$ and the
porosity $\phi_i$ at the position of the particle are needed.
These can be obtained by a linear interpolation of the fields $\grad P_j$
and $\phi_j$ from the gas algorithm:
\begin{equation}
\phi_i=\sum_j \phi_j F(\vec{r}_j-\vec{r}_i)\;, \qquad 
\grad P_i=\sum_j \grad P_j F(\vec{r}_j-\vec{r}_i)\;,
\end{equation}
where the sum is taken over all grid points.
Note that $\grad P_i$ is a continuous function of the particle position $\vec{r}_i$.
There are no discontinuities at all boundaries.

\subsection{Application to Plug Conveying}
This method was applied to study plug conveying in both vertical
\cite{A7:Strauss1} and horizontal \cite{A7:Strauss2} tubes.  
Plug conveying is a special
case of pneumatic conveying, where grains are driven
through pipes by air flow.
Plug conveying occurs when the flux of grains through the pipe is relatively
high.
Currently plug conveying is gaining importance in industry, because
it causes a lower product degradation 
and pipeline erosion than dilute phase conveying.

Unfortunately, current models~\cite{A7:KON0095,A7:SIE0096}
of plug conveying disagree
even on the prediction of such basic quantities
as the pressure drop and the total mass flow, and
these quantities have a great impact in industrial applications.
One of the reasons for the lack of valid models is that
it is difficult to study plugs experimentally in a detailed way.
Usually experimental setups are limited to the measurement of the
local pressure drop, the total mass flux and the velocity of plugs.
Simulational studies are handicapped by the high computational costs
for solving the gas flow and the particle-particle interaction,
and are therefore mostly limited to two dimensions.

Using the above-described method, we were able 
to provide a detailed view of plugs.
This approach provides access to important parameters like the porosity
and velocity of the granulate and the shear stress on the wall
at relatively low computational costs.
Contrary to the experiments,
it is possible to access these parameters 
at high spatial resolution and without influencing the process of 
transportation at all. 
Additional to plug profiles, characteristic curves of the pressure drop 
and the influence of simulation parameters can be measured.

\subsection{Results}
\begin{figure}
\includegraphics[width=\textwidth]{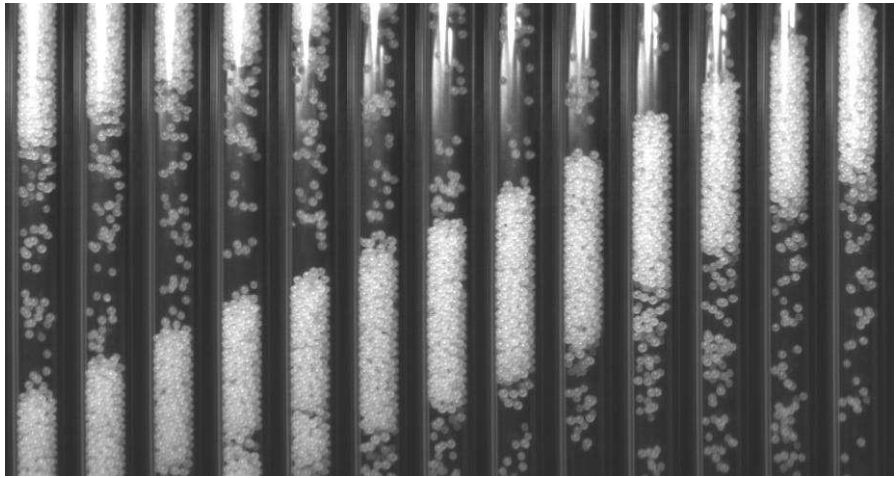}
\caption{\label{sean:experiment}
A series of photos showing a plug moving upwards.
The height of the shown tube is $9\,\mbox{cm}$, the frame~rate is $30\,\mbox{Hz}$}
\end{figure}

In Fig.\ \ref{sean:experiment}, we show a series of photos of plug conveying
taken by Karl Sommer and Gerhard Niederreiter of TU M\"unchen.
The particles are wax beads  of
diameter $d=1.41\,\mbox{mm}$, density $\varrho_s=937\,\mbox{kg/}\mbox{m}^3$ 
and a Coulomb coefficient of $0.21$.
The experimental transport
channel is a vertical tube (PMMA) of length $l=1.01\,\mbox{m}$ 
and of internal diameter $D_t=7\,\mbox{mm}$.
The air is injected at a constant flow rate of $2.2\,\ell/\mbox{min}$
at the bottom of the tube.  As one can easily see, grains travel in clusters
up the tube.

\begin{figure}
\includegraphics[width=\textwidth]{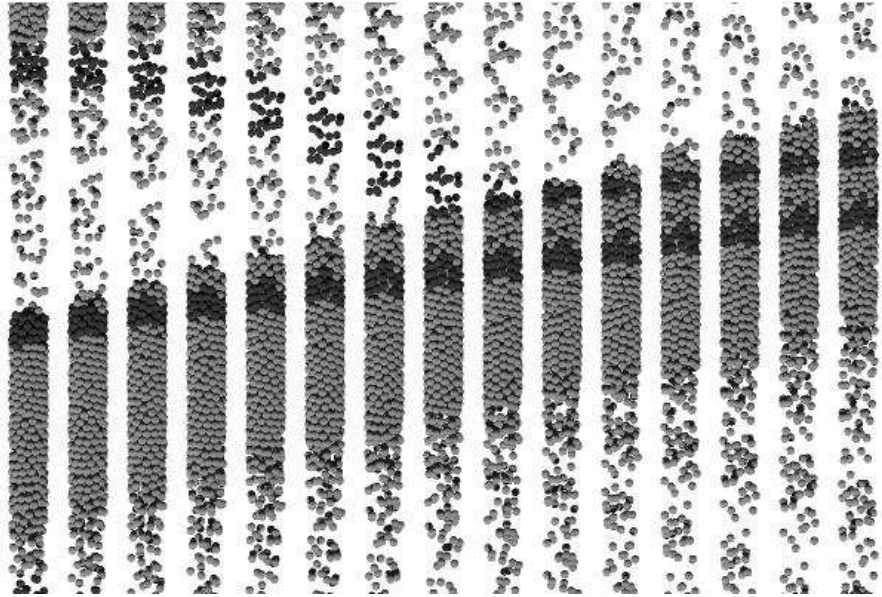}
\caption{\label{sean:simulation}
A series of simulation snapshots showing a plug moving upwards.
The height of the shown tube is $12\,\mbox{cm}$, the frame rate is $100\,\mbox{Hz}$}
\end{figure}

The simulations were carried out in a system that matched as closely as possible
the experimental one.
The same mode of transport was observed, as shown in
Fig.\ \ref{sean:simulation}.  Not only is there a qualitative resemblance
between Figs.\ \ref{sean:experiment} and \ref{sean:simulation}, but the
simulations give the same value for the pressure drop as the experiments.
The success of the model permitted a thorough study of the plugs to be carried
out.  For example, so-called ``characteristic curves'', where pressure drop
is displayed as a function of gas velocity, could be calculated.  The
simulations also allow the study of the effects of parameters
not easily controlled experimentally, such as the air viscosity and particle
friction.  The speed, density, size, and number of plugs were analyzed.  In 
addition, the detailed structure of plugs could be studied.  For example,
the variation of density, velocity, and different components of the
stress tensor were evaluated inside the plugs.  All this information should
help researchers to develop better models of plug conveying.

\section{Conclusion} \label{A7:Conclusions}
In this paper we have discussed the properties of various simulation
techniques for particles in fluids and demonstrated that there is no
perfect candidate that is able to simulate all systems of interest and
to utilize the available resources as efficient as possible. For each
individual problem, one has to choose the method of choice carefully:  
while stochastic rotation dynamics is well suited to simulate systems like
clay-like colloids where Brownian motion is important, the lattice
Boltzmann method is not able to resolve the stochastic motion of the
particles without modifications of the method. However, in cases where
thermodynamic fluctuations are neglectably small, this approach is much more
efficient than stochastic rotation dynamics. Like conventional
Navier-Stokes solvers, the fluid flow can be resolved in great detail, but
the lattice Boltzmann method is much easier to implement and to
parallelize. It is of particular advantage if complicated boundary
conditions like non-spherical particles or complex channel geometries come
into play. The implementation of Navier-Stokes solvers on the other hand
can be based on a long-standing and widespread experience with these
techniques allowing to create very efficient solvers.
In macroscopic systems like the movement of granular particles
in air, the exact properties of the flow field are not necessary to
understand experimentally observable parameters. Therefore,
computationally much less demanding techniques like a coarse-grained
description of the fluid should be applied. 

\section*{Acknowledgements}
We would like to thank all former members of the group who contributed to
the projects related to the simulation of particles in fluids.


\end{document}